\documentclass[twocolumn,showpacs,eqsecnum,pra,aps]{revtex4-1}

\usepackage{amsmath}
\usepackage{color}
\usepackage{amssymb}
\usepackage{epsfig}
\usepackage{soul}
\usepackage{color}

\begin{document}
\title{ Einstein-Podolsky-Rosen-like separability indicators \\
for two-mode Gaussian states }
\author{ Paulina Marian$^{1,2}$}
\email{paulina.marian@g.unibuc.ro}
\author{ Tudor A. Marian$^{1}$}
\email{tudor.marian@g.unibuc.ro}
\affiliation{ $^1$Centre for Advanced  Quantum Physics,
Department of Physics, University of Bucharest,
R-077125 M\u{a}gurele, Romania}
\affiliation{$^2$Department of Physical Chemistry, University of Bucharest,\\
Boulevard Regina Elisabeta 4-12, R-030018  Bucharest, Romania}


\begin{abstract}
We investigate the separability of the two-mode Gaussian states by using 
the variances of a pair of Einstein-Podolsky-Rosen (EPR)-like observables. 
Our starting point is inspired by the general necessary condition of separability introduced 
by Duan {\em et al.} [Phys. Rev. Lett. {\bf 84}, 2722 (2000)].  We evaluate the minima 
of the normalized forms of both the product and sum of such variances, 
as well as that of a regularized sum. Making use of Simon's separability criterion, 
which is based on the condition of positivity of the partial transpose (PPT) 
of the density matrix [Phys. Rev. Lett. {\bf 84}, 2726 (2000)], we prove that these minima 
are separability indicators in their own right. They appear to quantify the greatest amount 
of EPR-like correlations that can be created in a two-mode Gaussian state by means 
of local operations. Furthermore, we reconsider the EPR-like approach 
to the separability of two-mode Gaussian states which was developed by Duan {\em et al.} 
with no reference to the PPT condition. By optimizing the regularized form of their 
EPR-like uncertainty sum, we derive a separability indicator for any two-mode Gaussian state.  
We prove that the corresponding EPR-like condition of separability is manifestly equivalent 
to Simon's PPT one. The consistency of these two distinct approaches (EPR-like and PPT) 
affords a better understanding of the examined separability problem, whose explicit solution 
found long ago by Simon covers all situations of interest.
\end{abstract}
\maketitle

\section{Introduction}

Detecting and measuring quantum entanglement represent one of the goals 
of quantum information science. In the last two decades, a large amount of work 
has been invested in writing  efficient separability criteria for both discrete- and  
continuous-variable  systems. Although bipartite entanglement seems to be 
the simplest to detect and measure, practically we are still forced to apply 
different criteria when discussing it for mixed states of composite systems. 

As shown by Peres \cite{Peres}, a necessary condition of separability for an arbitrary 
two-party state is the requirement to have a non-negative partially 
transposed density matrix. In the case of discrete-variable systems, this requirement of 
positive partial transposition (PPT) is also a sufficient condition of separability only for states 
on ${\mathbb C}^2 \otimes {\mathbb C}^2$ and ${\mathbb C}^2 \otimes {\mathbb C}^3$
Hilbert spaces \cite{HHH}. For bipartite states of continuous-variable systems, 
the PPT condition was first applied by Simon \cite{Simon}. Specifically, Simon proved 
that preservation of the non-negativity of the density matrix under partial transposition  
is not only a necessary, but also a sufficient condition for the separability of two-mode 
Gaussian states (TMGSs).  Moreover, the partial transposition criterion 
could be expressed in an elegant symplectically invariant form valid for any TMGS \cite{Simon}.
To detect the continuous-variable entanglement for non-Gaussian states, Shchukin and Vogel 
have derived an infinite series of inequalities for the moments of the state required by the PPT 
condition \cite{SV}. Similar inequalities were obtained in Refs. \cite{BA2005,HZ2006}.

A somewhat parallel method to get general inseparability criteria for two-mode states 
originates in a practical procedure proposed by Reid for demonstrating 
the Einstein-Podolsky-Rosen (EPR) paradox \cite{EPR} in a non-degenerate parametric 
amplifier \cite{Reid1}. This was done by using  two non-local observables linearly built 
with the canonical quadrature operators of the modes, 
${\hat q}_j, {\hat p}_j, (j=1,2)$ \cite{Reid1,Reid2}:
\begin{align}
{\hat Q}(\lambda):= {\hat q}_1-\lambda {\hat q}_2,
\quad {\hat P}(\mu):= {\hat p}_1+\mu {\hat p}_2,
\label{Reid}
\end{align}
where $\lambda$ and $\mu$ are adjustable positive parameters.
As a consequence of their commutation relation, 
\begin{align} 
[{\hat Q}(\lambda), {\hat P}(\mu )]=i (1-\lambda \mu) {\hat I},  
\label{CR-Reid}
\end{align} 
we get the weak (Heisenberg) form of the uncertainty principle, 
\begin{align} 
\Delta Q(\lambda)\Delta  P(\mu) \geqq \frac{1}{2}|1-\lambda \mu|,
\label{prodR}
\end{align}
which has to be fulfilled by any quantum state. In Eq.\ (\ref{prodR}) and in the sequel as well, 
$(\Delta A)_{\hat \rho}$ denotes the standard deviation of the observable $\hat{A}$ in the
state $\hat{\rho}$, which is the square root of the variance 
$$\left[ (\Delta A)_{\hat \rho} \right]^2:={\left \langle \left( {\hat A}
-{\langle {\hat A}\rangle }_{\hat \rho}{\hat I}\right)^2 \right \rangle }_{\hat \rho}=
{\langle {\hat A}^2\rangle }_{\hat \rho}-\left( {\langle {\hat A} \rangle}_{\hat \rho} \right)^2.$$
Unless  $\lambda \mu=1$, the operators\ (\ref{Reid}) are not genuine EPR observables
since they do not commute. In Refs.\cite{Reid1,Reid2}, a possible experimental
observation of the inequality 
\begin{equation}
\Delta Q(\lambda)\Delta  P(\mu) < \frac{1}{2}
\label{EPR}
\end{equation} 
is interpreted as a sufficient condition to detect an EPR paradox. It is interesting 
to remark that the EPR paradox  \cite{EPR} as shown  in Eq.\ (\ref{EPR}) and the concept
of steering introduced by Schr\" odinger  \cite{Schr}  as analyzed in Refs.\cite{W1,W2,W3}  
were proven to be equivalent descriptions of non-locality. Moreover, the EPR-steering 
turned out to be a  different kind of non-locality stronger than quantum inseparability \cite{W1,W2}.

An important piece of progress in studying separability starting with the uncertainty principle was made 
by Duan {\em et al.} \cite{Duan}.  They have introduced another family of EPR-like uncertainties 
in terms of the variances of  the non-local one-parameter operators:
\begin{align}
& \hat Q(\alpha):=\alpha {\hat q}_1-\frac{1}{\alpha}{\hat q}_2,    \qquad 
\hat P_{\pm}(\alpha):=\alpha {\hat p}_1 \pm \frac{1}{\alpha}{\hat p}_2,     \qquad    
 (\alpha >0).
\label{Duan}
\end{align}
The commutation relations
\begin{align} 
[{\hat Q}(\alpha), {\hat P}_{\pm}(\alpha )]=i\left( {\alpha}^2 \mp \frac{1}{{\alpha}^2}\right) {\hat I} 
\label{CR-Reid1}
\end{align} 
lead to the product-form uncertainty relations for the EPR-like observables\ (\ref{Duan}),  
\begin{align} 
\Delta Q(\alpha)\Delta P_{\pm}(\alpha) \geqq \frac{1}{2}
\left| {\alpha}^2 \mp \frac{1}{{\alpha}^2}\right|,
\label{prodD}
\end{align}
which imply the sum-form inequalities:
\begin{align}
\left[ \Delta Q(\alpha)\right]^2+\left[ \Delta P_{\pm}(\alpha)\right]^2
\geqq \left| {\alpha}^2 \mp \frac{1}{{\alpha}^2}\right|.
\label{sumD}
\end{align}
In Ref.\cite{Duan}, the inequalities\ (\ref{sumD}) are strengthen for separable states:
a necessary condition of separability, consisting of one-parameter family of inequalities, 
is thereby established for any two-mode state. Moreover, for the special class of TMGSs, 
the strongest of these inequalities is proven to be also a sufficient condition of separability.
Some other necessary conditions of separability which employ pairs of more general 
non-local observables depending on one or more parameters have been pointed out 
\cite{Simon,MGVT,GMVT}. They are expressed in terms of the product 
or the sum of variances of such EPR-like observables. 

The aim of this paper is twofold. On the one hand, we tackle a characterization 
of the separability of TMGSs based on the analysis of EPR-like correlations.
We write full criteria of separability for TMGSs in terms of both the product 
and the sum of variances of two EPR-like observables. Their formulation requires 
an explicit use of Simon's PPT criterion of separability. On the other hand, 
we offer a fresh look at the work of Duan {\em et al.} \cite{Duan}, which is based 
on an EPR-like uncertainty of sum form and is fully independent of the PPT approach.
First, a suitable optimization of this EPR-like uncertainty sum leads eventually 
to an EPR-like indicator of separability for TMGSs, which is directly related 
to a specific classicality condition in the quantum optical sense. Second, we prove 
that the corresponding condition of separability is manifestly equivalent to that obtained 
by Simon in the framework of the PPT approach. This is  an important aspect 
of the Gaussian separability problem which remained unclear for many years 
and was falsely doubted quite recently \cite{K2009,K2012}.

The article is structured in the following way. In Sec. II, we recall the EPR-like 
necessary conditions of separability for two-mode states, as discussed 
by Giovannetti {\em et al.} in Ref. \cite{GMVT}. Section III is an overview 
of several useful properties of the undisplaced TMGSs. In Sec. IV, the product function 
occurring in Eq.\ (\ref{prodR}) and one of the sum functions from Eq.\ (\ref{sumD}) 
are normalized according to the formulae presented in Sec. II. 
Here we evaluate their minimal values over the whole range 
of the parameters  defining the EPR-like observables, as well as under local unitary actions. 
We thus find the maximal EPR-like correlations over the family of equally entangled states.
A two-parameter regularized sum function is examined along the same lines 
in Sec. V,  where its minimum is found. On the one hand, the formulae established 
in Secs. IV and V allow us to derive the Peres-Simon PPT necessary condition 
of separability for a TMGS from each of the three EPR-like necessary conditions 
we have employed. On the other hand, they enable us to formulate full criteria 
of separability for TMGSs, owing to the PPT criterion of separability\cite{Simon}. 
In Subsec. VI A, we describe the optimization of the regularized EPR-like 
uncertainty sum introduced by Duan {\em et al.} \cite{Duan}. This leads to the so-called 
standard form II of the covariance matrix (CM), whose existence is proven in Subsec. VI B. 
Then, in Subsec. VI C, we build an EPR-like separability indicator in terms of the standard form II
of the CM. Subsection VI D is devoted to some special cases when the EPR-like 
separability indicator can be determined explicitly. In Sec. VII, we establish directly 
the equivalence of the EPR-like and PPT conditions of separability. 
Section  VIII summarizes our main results insisting on the connection between 
the EPR-like and PPT separability conditions for a TMGS. Appendix A presents 
some nontrivial identities involving the standard-form parameters of the CM 
of a TMGS. In Appendix B, we prove the positivity of the Hessian matrices 
of the three EPR-like correlation functions examined in Secs. IV and V, 
when evaluated at their stationary points.

\section{Separable two-mode states} 

Let us consider a pair of EPR-like observables which are linear combinations 
of the canonical quadrature operators of the two modes:
\begin{align}
& \hat Q:={\alpha}_1 {\hat q}_1- {\alpha}_2 {\hat q}_2,  \qquad
\hat P:={\beta}_1 {\hat p}_1+{\beta}_2 {\hat p}_2,  \nonumber \\
& ( {\alpha}_j >0, \;\;  {\beta}_j >0: \;  j=1,2 ).
\label{GMVT}
\end{align}
The coordinates and momenta in Eq.\ (\ref{GMVT}) are defined in terms of the amplitude
operators of the modes:
\begin{align}
& \hat q_j:=\frac{1}{\sqrt{2}}\left( \hat a_j+\hat a_j^{\dag}\right),  \qquad 
\hat p_j:=\frac{1}{\sqrt{2}i}\left( \hat a_j-\hat a_j^{\dag}\right),   \nonumber \\
& (j=1,2 ).
\label{quad}
\end{align}
Obviously, Reid's pair of one-parameter observables built with independent 
parameters\ (\ref{Reid}), as well as the single-parameter one\ (\ref{Duan}) 
are particular cases of EPR-like operators\ (\ref{GMVT}). The commutation relation
\begin{align} 
[{\hat Q}, {\hat P}]=i\left( {\alpha}_1{\beta}_1-{\alpha}_2{\beta}_2 \right) {\hat I} 
\label{CR-GMVT}
\end{align} 
shows that the operators\ (\ref{GMVT}) are genuine  EPR observables if and only if
${\alpha}_1{\beta}_1= {\alpha}_2{\beta}_2.$ The corresponding Heisenberg
uncertainty relation, 
\begin{align} 
{\Delta Q}\, {\Delta P} \geqq \frac{1}{2}\left| {\alpha}_1{\beta}_1-{\alpha}_2{\beta}_2\right|,
\label{prodG}
\end{align}
entails the sum-type inequality
\begin{align}
\left( \Delta Q \right)^2+\left( \Delta P\right)^2
\geqq \left| {\alpha}_1{\beta}_1-{\alpha}_2{\beta}_2\right|.
\label{sumG}
\end{align}

If the two-mode state is separable, i. e., it is a convex combination of product states, 
$$\hat \rho_s:=\sum_{k} w_k {\hat \rho}_1^{(k)}\otimes {\hat \rho}_2^{(k)}, \qquad  
\left( w_k >0,  \quad  \sum_{k} w_k=1 \right),$$
then the product $(\Delta Q)_s(\Delta P)_s$ has a stronger lower bound 
than in Eq.\ (\ref{prodG}) \cite{GMVT}:
\begin{align} 
(\Delta Q)_s(\Delta P)_s \geqq \frac{1}{2}\left( {\alpha}_1{\beta}_1+{\alpha}_2{\beta}_2\right).
\label{prodsG}
\end{align}
Accordingly, for the sum of variances, an inequality stronger than Eq.\ (\ref{sumG}) holds:
\begin{align}
\left[ (\Delta Q)_s \right]^2+\left[ (\Delta P)_s\right]^2
\geqq {\alpha}_1{\beta}_1+{\alpha}_2{\beta}_2.
\label{sumsG}
\end{align}
It is useful to specialize the necessary conditions for the separability of a two-mode state,
Eqs.\ (\ref{prodsG}) and\ (\ref{sumsG}), to the pairs of EPR-like observables\ (\ref{Reid})
and\ (\ref{Duan}). We get the following two sets of inequalities: 
\begin{align} 
[\Delta Q(\lambda)]_s [\Delta P(\mu)]_s \geqq \frac{1}{2}\left( 1+\lambda \mu \right),
\label{prodsR}
\end{align}
\begin{align}
\left\{ [\Delta Q(\lambda)]_s \right\}^2+\left\{ [\Delta P(\mu)]_s \right\}^2
\geqq 1+\lambda \mu \, ; 
\label{sumsR}
\end{align}
\begin{align} 
[\Delta Q(\alpha)]_s [\Delta P_{\pm}(\alpha)]_s \geqq \frac{1}{2}\left(  {\alpha}^2
+\frac{1}{{\alpha}^2} \right),
\label{prodsD}
\end{align}
\begin{align}
\left\{ [\Delta Q(\alpha)]_s \right\}^2+\left\{ [\Delta P_{\pm}(\alpha)]_s \right\}^2
\geqq {\alpha}^2+\frac{1}{{\alpha}^2}.
\label{sumsD}
\end{align}
The necessary conditions of separability\ (\ref{sumsD}) have first been derived 
in Ref.  \cite{Duan}.Following a similar pattern, the product-form 
conditions\ (\ref{prodsR}) and \ (\ref{prodsD}) have then been written 
in Refs.\cite{MGVT,GMVT}. Interestingly enough, it was shown in Ref.\cite{GMVT} 
that the product-form necessary conditions of separability are stronger than the corresponding
sum-form ones. We also remark that some particular forms of the necessary conditions 
of separability\ (\ref{prodsR})--\ (\ref{sumsD})  have been used to detect entanglement 
in experiments \cite{Bowen,Josse,Buono}. 

\section{Two-mode Gaussian states}

For a long time, the Gaussian states (GSs) of the quantum radiation field are known to be 
of central importance in various areas of quantum optics. In general, the GSs play 
an important role for those quantum systems involving a quadratic bosonic Hamiltonian 
that generates correlations between bosonic modes. They are achieved in condensed matter,
as well as in atomic ensembles such as trapped ions or Bose-Einstein condensates. 
The GSs of light are also largely employed in quantum information processing
with continuous variables; their usefulness has been reviewed 
in Refs. \cite{EP,BL,QC3}.

Especially accessible and insightful are the TMGSs. They have a privileged position 
among the bipartite quantum states of continuous-variable systems. Specifically, 
their properties are fully determined by the first- and second-order moments 
of the quadrature observables of the modes \cite{PTH2001}.  

From the theoretical point of view, the TMGSs represent a perfect test bed 
for studying entanglement or other kinds of correlations between 
the modes \cite{A2016}. Note that the entanglement of a TMGS depends only on its CM, 
so that it can be checked and quantified easier than for other two-party states.
Here we recollect just the strictly necessary notions 
and results \cite{PTH2001,PTH2003,ASI} that enable us to discuss separability issues. 
Since these are not affected by translations in the phase space, it is sufficient 
to deal with undisplaced (zero-mean) TMGSs.

Recall that the characteristic function (CF) of an undisplaced TMGS $\hat \rho$
is a real exponential:
\begin{equation}
\chi (u)=\exp\left[ -\frac{1}{2}(Ju)^{T}{\mathcal V}(Ju) \right].
\label{CF}
\end{equation}
Its argument is a dimensionless vector $u \in {\mathbb R}^4$ whose components
are  eigenvalues of the canonical quadrature operators of the modes:
\begin{equation}
u^T:=(q_1,\; p_1,\; q_2,\; p_2).
\label{uT}
\end{equation}
$J$ designates the standard matrix of the symplectic form on ${\mathbb R}^4$,
which is block-diagonal and skew-symmetric:
\begin{align}
J:=J_1 \oplus J_2:   \qquad   J_1=J_2:=\left( 
\begin{matrix}
0  & 1\\ -1 & 0
\end{matrix}
\right).
\label{J}
\end{align}
The TMGS $\hat \rho$ is entirely specified, via the CF\ (\ref{CF}), by the real 
and symmetric $4 \times 4$ CM, which is denoted ${\mathcal V}$. Its entries 
are expectation values of products of the deviations from the means 
of the quadratures\ (\ref{quad}). The order of the rows and columns 
is indicated by the current vector\ (\ref{uT}). The CM ${\mathcal V}$ fulfills 
the strong form (Robertson-Schr\"odinger) of the uncertainty relations 
for the canonical quadrature observables\ (\ref{quad}):  
\begin{equation}
{\mathcal V}+\frac{i}{2}J \geqslant 0.
\label{RS}
\end{equation}
The above requirement that the complex matrix  ${\mathcal V}+\frac{i}{2}J$ 
has to be positive semidefinite is a necessary and sufficient condition 
for the existence of the GS ${\hat \rho}$ \cite{Simon}. It implies the inequality
\begin{equation}
{\cal D}:=\det \left( {\mathcal V}+\frac{i}{2}J \right) \geqq 0,
\label{D}
\end{equation}
as well as the general property that the CM ${\mathcal V}$ is positive definite: 
${\mathcal V}>0$. By contrast, the saturation equality ${\cal D}=0$ 
is a quite special feature. However, it is shared by all the pure GSs 
and also by an interesting class of mixed ones. All these states are said 
to be at the physicality edge.

 It is often convenient to partition the CM ${\mathcal V}$ into $2\times 2$ submatrices:
\begin{align}
{\mathcal V}=\left(
\begin{matrix} 
\mathcal V_{1}\;   &   \; {\mathcal C} \\   \\
{\mathcal C}^ T \; &   \;  \mathcal V_{2} 
\end{matrix}
\right).
\label{part}
\end{align} 
Here $\mathcal V_{j}, \;  (j=1,\, 2)$,  denote the CMs of the individual single-mode 
reduced GSs, while $ {\mathcal C}$ displays the cross-correlations between the modes.

Let us mention that a symplectic transformation $S$ of the canonical quadrature 
observables\ (\ref{quad}) in the Heisenberg picture induces a congruence 
transformation of any CM:
\begin{equation}
{\mathcal V}^{\prime}=S\, {\mathcal V}\, S^T,   \qquad    S \in {\rm Sp}(4,\mathbb{R}). 
\label{congr}
\end{equation}
At the same time, a unitary operator ${\hat U}(S)$ acting on the two-mode Fock space 
${\mathcal H}_1 \otimes {\mathcal H}_2$ is associated to the symplectic matrix $S$. 
Remarkably, the corresponding transformation of the TMGS in the Schr\"odinger picture,
${\hat \rho}^{\prime}={\hat U}(S)\, {\hat \rho}\, {\hat U}^{\dag}(S)$, preserves the
Gaussian nature of the state. Note that, if the symplectic transformation $S$ consists
of two separate single-mode ones,
\begin{equation} 
S=S_1 \oplus  S_2:  \qquad   S\in{\rm Sp}(2,\mathbb{R})\times {\rm Sp}(2,\mathbb{R}),
\label{local}
\end{equation}
then the above transformation of the TMGS ${\hat \rho}$ does not affect its amount 
of entanglement. It has been proven in Ref. \cite{Duan} and just exploited 
in Ref. \cite{Simon} the existence of a local symplectic transformation\ (\ref{local}) 
that leads to a standard form of the CM:
\begin{align}
{\mathcal V}(1,1):=\left(
\begin{matrix} 
b_1\;   & \;  0\;  & \;  c\;  & \;  0  \\  \\
0\;   & \;  b_1\;  & \;  0\;  & \;  d   \\  \\
c\;  & \;  0\;  & \;  b_2\;   & \;  0   \\  \\
0\;  & \;  d\;  & \;  0\;  &  \;  b_2   
\end{matrix}
\right),
\label{standard}
\end{align} 
depending on four parameters $b_1,\, b_2,\, c,\, d$. With no loss of generality,
one can choose $b_1 \geqq b_2>0$ and $c \geqq |d|$. Following 
Refs. \cite{Duan,Simon}, we apply two independent one-mode squeeze 
transformations to the standard-form CM\ (\ref{standard}). Let us denote 
the corresponding scaling factors by
\begin{equation} 
u_1:=e^{2r_1}\geqq 1,  \qquad    u_2:=e^{2r_2}\geqq 1,
\label{scal}
\end{equation} 
where $r_1$ and $r_2$ are the squeeze parameters in the two modes.
The transformed state has a {\em scaled standard-form} CM with the block 
structure\ (\ref{part}):
\begin{align}
{\mathcal V}(u_1,\, u_2) =\left(
\begin{matrix} 
{\mathcal V}_1(u_1)\;  &  \;  {\mathcal C}\left( \sqrt{u_1u_2} \right)  \\   \\
{\mathcal C}\left( \sqrt{u_1u_2} \right) \;  &  \;  {\mathcal V}_2(u_2) 
\end{matrix}
\right).
\label{ssfCM}
\end{align} 
All its $2\times 2$ submatrices are diagonal. The CMs of the single-mode 
reduced GSs read
\begin{align}
{\mathcal V}_j(u_j)=\left(
\begin{matrix} 
b_j\,u_j\;   &   \;  0  \\   \\
0\;   &   \;  \frac{b_j}{u_j}  
\end{matrix}
\right),    \qquad  (j=1,2),
\label{V_j}
\end{align}
while the cross-correlation matrix is
\begin{align}
{\mathcal C}\left( \sqrt{u_1u_2} \right)=\left(
\begin{matrix} 
c\sqrt{u_1u_2} \;   &   \;  0  \\   \\
0\;   &   \;   \frac{d}{\sqrt{u_1u_2}}
\end{matrix}
\right).
\label{C}
\end{align}
The TMGSs whose CMs have a scaled standard form\ (\ref{ssfCM})-\ (\ref{C})
constitute a class characterized  by the set of four standard-form parameters  
$\{ \, b_1,\, b_2,\, c,\, d \} $ of the given TMGS ${\hat \rho}$, which are
${\rm Sp}(2,\mathbb{R})\times {\rm Sp}(2,\mathbb{R})$-invariant. The states
of this  class are labeled by the pair of local squeeze factors $\{u_1,\, u_2\}$
\cite{Duan}. They are locally unitary similar to the given TMGS ${\hat \rho}$ 
and thereby possess precisely its amount of entanglement. 

The Robertson-Schr\"odinger uncertainty relation\ (\ref{RS}) reduces 
to the following restrictions for the standard-form parameters: 
\begin{align}
b_1 \geqq \frac{1}{2},   \qquad     b_2 \geqq \frac{1}{2};
\label{GS1}
\end{align}
\begin{align} 
& b_1 b_2-c^2 \geqq \frac{1}{4} \max \left\{ \frac{b_1}{b_2}, \;
\frac{b_2}{b_1} \right\}  \geqq \frac{1}{4},     \notag \\  
& b_1 b_2-d^2 \geqq \frac{1}{4} \max \left\{ \frac{b_1}{b_2}, \;
\frac{b_2}{b_1} \right\}  \geqq \frac{1}{4};  
\label{GS2}
\end{align}
\begin{align}
& {\mathcal D}=\left( b_1 b_2-c^2 \right) \left( b_1 b_2-d^2 \right)     \notag \\
& -\frac{1}{4} \left( b_1^2 +b_2^2+2cd \right) +\frac{1}{16} \geqq 0.
\label{GS3}
\end{align}
The inequalities\ (\ref{GS2}) give rise to another one involving the symplectic invariant
$\det({\mathcal V})=\left( b_1 b_2-c^2 \right) \left( b_1 b_2-d^2 \right) $ and thus 
concerning the purity of the state:
\begin{align}
\det({\mathcal V}) \geqq \frac{1}{16}  \quad  \Longleftrightarrow     \quad 
{\rm Tr}\left( {\hat \rho}^2 \right)   \leqq 1.
\label{purity}
\end{align}

According to Williamson's theorem \cite{W}, the CM ${\mathcal V}$  of any TMGS 
${\hat \rho}$ is congruent to a diagonal CM via a symplectic matrix\ (\ref{congr}), which
is unique up to the sign. The corresponding  diagonal entries, denoted 
${\kappa}_{\pm},\,$ are positive, each one occurring twice. They are called 
the symplectic eigenvalues of the CM  ${\mathcal V}$ \cite{VW}. By virtue of Eq.\ (\ref{congr}), 
the symplectic invariants $\det({\mathcal V})$ and ${\mathcal D}$ factor as follows:
\begin{align}
\det({\mathcal V})=\left( {\kappa}_{+} \right)^2 \left( {\kappa}_{-} \right)^2, 
\label{detV1}
\end{align}
\begin{align}
{\mathcal D}=\left[ \left( {\kappa}_{+} \right)^2-\frac{1}{4} \right]
\left[ \left( {\kappa}_{-} \right)^2-\frac{1}{4} \right] \geqq 0,
\label{D1}
\end{align}
with
\begin{align}
{\kappa}_{+} \geqq {\kappa}_{-} \geqq \frac{1}{2}. 
\label{kappa}
\end{align} 
We employ Eqs.\ (\ref{detV1}) and\ (\ref{D1}) to get the symplectic eigenvalues in terms 
of the standard-form parameters \cite{VW}:
\begin{align}
& \left( {\kappa}_{\pm} \right)^2=\frac{1}{2}\left[ \left(b_1^2+b_2^2 +2cd \right) 
\pm \sqrt{\Delta} \right],      \notag \\
& \Delta:=\left( b_1^2+b_2^2+2cd \right)^2 -4\, \det({\mathcal V})    \notag \\ 
& =\left( b_1^2-b_2^2 \right)^2+4\left( b_1 c+b_2 d \right) \left( b_2 c+b_1 d \right) 
\geqq 0. 
\label{kappa+-}  
\end{align}

In order to formulate Simon's separability criterion for TMGSs, we have to consider 
the partial transpose ${\hat \rho}^{\rm PT}$  of the TMGS ${\hat \rho}$. Basically, partial 
transposition preserves the Gaussian character of the operator, does not modify 
the standard-form parameters $\, b_1,\, b_2,\, c,\, $ and changes the sign of $d$. 
Therefore, ${\hat \rho}^{\rm PT}$ is an undisplaced two-mode Gaussian operator whose CM  
${\mathcal V}^{\rm PT}$ has the standard-form parameters  $\{ \, b_1,\, b_2,\, c,\, -d \} $ 
\cite{Simon}. First, Simon has proven a lemma asserting that any TMGS with $d \geqq 0$ 
is separable. Then, he has shown that Peres' necessary condition of separability, 
which claims that ${\hat \rho}^{\rm PT}$ should be a quantum state, i. e.,
\begin{equation}
{\mathcal V}^{\rm PT}+\frac{i}{2}J \geqslant 0,
\label{Peres}
\end{equation}
is also a {\em sufficient} one \cite{Simon}.  The condition\ (\ref{Peres})  
for the existence of a GS ${\hat \rho}^{\rm PT}$ reduces to the inequality
\begin{equation}
{\mathcal D}^{\rm PT}:=\det \left( {\mathcal V}^{\rm PT}+\frac{i}{2}J \right) \geqq 0,
\label{DPT}
\end{equation}
which is Simon's criterion of separability. It is usually written in the form\ (\ref{GS3})
with $d \to -d$. Therefore, a TMGS is separable if and only if  the following inequality 
is fulfilled:
\begin{align}
{\mathcal D}^{\rm PT}=\det({\mathcal V})-\frac{1}{4}\left( b_1^2+b_2^2-2 c d \right)
+\frac{1}{16} \geqq 0.
\label{DPT1} 
\end{align}
Accordingly, the identity
\begin{align}
{\mathcal D}^{\rm PT}={\mathcal D}+cd
\label{d>0} 
\end{align}
displays the property that all the TMGSs with $d \geqq 0$ are separable, 
in agreement with Simon's lemma  \cite{Simon}.

\section{Normalized separability indicators}

We write the variances of the EPR-like observables\ (\ref{GMVT}) for a TMGS  
whose CM has a scaled standard form\ (\ref{ssfCM})--\ (\ref{C}): 
\begin{align}
\left( \Delta Q \right)^2={\alpha}_1^2\, b_1 u_1+{\alpha}_2^2\, b_2 u_2
-2{\alpha}_1{\alpha}_2 c \sqrt{u_1u_2}\, ,
\label{varQ} 
\end{align}
\begin{align}
\left( \Delta P \right)^2={\beta}_1^2\, \frac{b_1}{u_1}+{\beta}_2^2\, \frac{b_2}{u_2}
+2{\beta}_1{\beta}_2 \frac{d}{\sqrt{u_1u_2}}.
\label{varP} 
\end{align}
Remark that, while all the TMGSs\ (\ref{ssfCM})--\ (\ref{C}) belonging to a class 
with fixed standard-form parameters possess the same amount of entanglement, 
their EPR-like variances\ (\ref{varQ}) and\ (\ref{varP}) depend, in addition, 
on the local squeezings $u_1, u_2$. In what follows, we consider two functions  
which are normalized as suggested by the separability 
lower bounds\ (\ref{prodsG}) and\ (\ref{sumsG}): 
\begin{align}
E\left( {\alpha}_1\, {\beta}_1,\, {\alpha}_2,\, {\beta}_2,\, u_1,\, u_2 \right) 
:=\frac{\left( \Delta Q \right)^2 \left( \Delta P \right)^2}
{\left( {\alpha}_1{\beta}_1+{\alpha}_2{\beta}_2\right)^2},
\label{E1} 
\end{align}
\begin{align}
F\left( {\alpha}_1,\, {\beta}_1,\, {\alpha}_2,\, {\beta}_2,\, u_1,\, u_2 \right) 
:=\frac{\left( \Delta Q \right)^2+\left( \Delta P \right)^2}
{{\alpha}_1{\beta}_1+{\alpha}_2{\beta}_2}.
\label{f1} 
\end{align}
According to Eqs.\ (\ref{prodsG}) and\ (\ref{sumsG}), for any separable
two-mode state (Gaussian or non-Gaussian), the following inequalities
are satisfied:  
\begin{align}
E \geqq \frac{1}{4},   \qquad    F \geqq 1. 
\label{EF} 
\end{align}
In the sequel, we develop a two-step program. First, in order 
to handle simpler functions, we diminish the number of their independent variables 
as much as possible by making suitable substitutions or choices. Second, 
we find the absolute minima of the resulting simpler functions, which prove to be 
manifest separability indicators. It is sufficient to restrict our search in this step 
to values $d<0$ in the variance\ (\ref{varP}).

\subsection{Uncertainty product}

We start by absorbing the scaling factors in Eqs.\ (\ref{varQ}) and\ (\ref{varP}) 
into four new positive parameters,
\begin{align}
{\alpha}_j^{\prime}= {\alpha}_j\sqrt{u_j},    \quad        
{\beta}_j^{\prime}= {\beta}_j \frac{1}{\sqrt{u_j}},     \quad   (j=1,2),
\label{prime} 
\end{align}
so that the variances become:
\begin{align}
\left( \Delta Q \right)^2= b_1\left( {\alpha}_1^{\prime} \right)^2
+b_2\left( {\alpha}_2^{\prime} \right)^2 -2c\,{\alpha}_1^{\prime}{\alpha}_2^{\prime}, 
\label{varQ1} 
\end{align}
\begin{align}
\left( \Delta P \right)^2=b_1\left( {\beta}_1^{\prime} \right)^2
+b_2\left( {\beta}_2^{\prime} \right)^2 +2d\,{\beta}_1^{\prime}{\beta}_2^{\prime}.
\label{varP1} 
\end{align}
Note the identity
\begin{align}
{\alpha}_1{\beta}_1+{\alpha}_2{\beta}_2={\alpha}_1^{\prime}{\beta}_1^{\prime}
+{\alpha}_2^{\prime}{\beta}_2^{\prime}.
\label{denom} 
\end{align}
After replacing Eqs.\ (\ref{varQ1})-\ (\ref{denom}) into Eq.\ (\ref{E1}), 
the function $E\left( {\alpha}_1^{\prime}\, {\beta}_1^{\prime},\, {\alpha}_2^{\prime},\, 
{\beta}_2^{\prime} \right) $ can be further simplified. As a matter of fact, it is
a function of two positive variables,
\begin{align}
\lambda:=\frac{{\alpha}_2^{\prime}}{{\alpha}_1^{\prime}},   \qquad 
\mu:=\frac{{\beta}_2^{\prime}}{{\beta}_1^{\prime}},
\label{lam-mu} 
\end{align}
which reads:
\begin{align}
E\left( \lambda,\, \mu \right) :=\frac{\left[ \Delta Q (\lambda) \right]^2 
\left[ \Delta P(\mu) \right]^2}{\left( 1+ \lambda \mu \right)^2}.
\label{E3} 
\end{align}
The numerator of the fraction in the r. h. s. of Eq.\ (\ref{E3}) is equal to the product
of the variances of Reid's EPR-like observables\ (\ref{Reid}) in a standard-form 
TMGS\ (\ref{standard}):
\begin{align}
\left[ \Delta Q (\lambda) \right]^2 =b_1 +b_2{\lambda}^2 -2c \lambda,
\label{varQ2} 
\end{align}
\begin{align}
\left[ \Delta P(\mu) \right]^2 =b_1 +b_2{\mu}^2 +2d \mu. 
\label{varP2} 
\end{align}
We prove the following statement.

{\bf Theorem 1.} {\em For any TMGS with $d<0$ the positive function 
$E\left( \lambda,\, \mu \right)$ has an absolute minimum equal to 
the square of the smallest symplectic eigenvalue ${\kappa}_{-}^{\rm PT}:$}
\begin{align}
E_m:=\min_{ \{ \lambda,\, \mu \} } E\left( \lambda,\, \mu  \right) 
=\left( {\kappa}_{-}^{\rm PT} \right)^2.  
\label{minE}
\end{align}
{\em Proof}. The first-order derivatives 
\begin{align}
& \frac{\partial \ln(E)}{\partial \lambda}=\frac{2 \left( b_2{\lambda}-c \right) }
{b_1 +b_2{\lambda}^2 -2c \lambda}-\frac{2\mu}{1+ \lambda \mu},      \notag \\
& \frac{\partial \ln(E)}{\partial \mu}=\frac{2 \left( b_2{\mu}+d \right) }
{b_1 +b_2{\mu}^2 +2d \mu}-\frac{2\lambda}{1+ \lambda \mu}
\label{der1}
\end{align}
vanish at a stationary point of the function\ (\ref{E3}). The resulting equations, 
\begin{align}
& \left[ \Delta Q (\lambda) \right]^2 =\frac{1}{\mu} \left( b_2{\lambda}-c \right)
\left( 1+ \lambda \mu \right),     \notag\\
& \left[ \Delta P(\mu) \right]^2 =\frac{1}{\lambda} \left( b_2{\mu}+d \right)
\left( 1+ \lambda \mu \right), 
\label{stat} 
\end{align}
have a unique solution:
\begin{align}
& {\lambda}_m=\frac{\left( b_1^2-b_2^2 \right)+\sqrt{{\Delta}^{\rm PT}}}
{2\left( b_1 c -b_2 d \right)},     \quad
{\mu}_m=\frac{\left( b_1^2-b_2^2 \right)+\sqrt{{\Delta}^{\rm PT}}}
{2\left( b_2 c -b_1 d \right)},
\label{minp}
\end{align}
where ${\Delta}^{\rm PT}$ is the discriminant in Eq.\ (\ref{A5}). In Appendix B
we have proven that the value\ (\ref{B4}),  
\begin{align}
E_m=\frac{  \left( b_2{\lambda}_m-c \right)\left( b_2{\mu}_m+d \right)} { {\lambda}_m {\mu}_m},
\label{E_m} 
\end{align}
is the absolute minimum of the function $E\left( \lambda,\, \mu \right) .$
By use of Eqs.\ (\ref{minp}) and\ (\ref{A5}), we find the formula
\begin{align}
E_m=\left( {\kappa}_{-}^{\rm PT} \right)^2.
\label{Emin}
\end{align} 
Therefore, the minimum of the normalized product\ (\ref{E3}) of two EPR-like uncertainties 
is equal to the square of the smallest symplectic eigenvalue ${\kappa}_{-}^{\rm PT}$. 
This concludes the proof. Note that although stated for $d<0$, Eq.\ (\ref{Emin}) still holds
for $0 \leqq d < \frac{b_2}{b_1}c$.

A previous relationship between ${\kappa}_{-}^{\rm PT}$  
and a product of EPR-like uncertainties was found by a different treatment 
and from another perspective in Ref. \cite{Poon-Law}.The normalized product 
of the EPR-like uncertainties\ (\ref{varQ2}) and \ (\ref{varP2}) was recently examined 
as a signature of entanglement for special families of TMGSs \cite{He1,He2,He3}. 
The present optimization approach was successfully applied in Ref. \cite{He3} to the class of 
the squeezed thermal states (STSs): $c=-d>0$ \cite{PTH2003}.

An interesting related problem for TMGSs regarding steerability criteria was recently 
put forward by Kogias and Adesso \cite{KA}. These authors have considered 
the product of EPR-like uncertainties occurring in the Reid condition\ (\ref{EPR}) 
and have  minimized it with respect to the pair of parameters $\{ \lambda,\mu \}$ 
and the local variables of a TMGS. They have thus recovered the explicit 
symplectically invariant formula for the condition of steerability of GSs 
by Gaussian measurements which has first been written in Refs. \cite{W1,W2}. 

\subsection{Uncertainty sum}

We find it convenient to simplify the function\ (\ref{f1}) by setting 
${\alpha}_1={\beta}_1=:\alpha,\, {\alpha}_2={\beta}_2=:\frac{1}{\alpha}:$
\begin{align}
F\left( {\alpha}^2,\, u_1,\, u_2 \right) :=\frac{ \left[ \Delta Q(\alpha) \right]^2 
+\left[ \Delta P_{+}(\alpha) \right]^2 }{ {\alpha}^2 +\frac{1}{{\alpha}^2} }.
\label{F2} 
\end{align}
The correlation function\ (\ref{F2}) is built with the variances of a pair of non-local 
one-parameter observables\ (\ref{Duan}) introduced 
in Ref. \cite{Duan}: 
\begin{align}
\left[ \Delta Q(\alpha) \right]^2=b_1 u_1 {\alpha}^2 
+b_2 u_2 \frac{1}{ {\alpha}^2 } -2c\sqrt{u_1\, u_2},
\label{varQ4} 
\end{align}
\begin{align}
\left[ \Delta P_{\pm}(\alpha) \right]^2 =\frac{b_1}{u_1} {\alpha}^2 
+\frac{b_2}{u_2} \frac{1}{ {\alpha}^2 } \pm \frac{2d}{ \sqrt{u_1\, u_2} }.
\label{varP4} 
\end{align}
Therefore, it reads:
\begin{align} 
& F\left( {\alpha}^2,\, u_1,\, u_2 \right) =\frac{1}{ {\alpha}^4 +1}  
\left[ b_1 \left( u_1+ \frac{1}{u_1} \right) {\alpha}^4  
-2\left( c\sqrt{u_1\, u_2}   \right. \right.   \notag \\
&  \left. \left.  -d \frac{1}{ \sqrt{u_1\, u_2} }\right) {\alpha}^2
+b_2 \left( u_2+ \frac{1}{u_2} \right) \right].
\label{F3} 
\end{align}
We state

{\bf Theorem 2.} {\em For any TMGS with $d<0$ the positive function 
$F\left( {\alpha}^2,\, u_1,\, u_2 \right)$ has an absolute minimum which is twice 
the smallest symplectic eigenvalue ${\kappa}_{-}^{\rm PT}:$} 
\begin{align}
F_m:=\min_{ \{ {\alpha}^2,\, u_1,\, u_2  \} } F\left( {\alpha}^2,\, u_1,\, u_2 \right)
=2{\kappa}_{-}^{\rm PT}.
\label{minF}
\end{align}
{\em Proof}. The first-order derivatives
\begin{align}  
& \frac{\partial F}{\partial \left( {\alpha}^2 \right)} =\frac{2}{\left( {\alpha}^4 +1 \right)^2 }  
\left\{ \left( c\sqrt{u_1\, u_2} -\frac{d}{ \sqrt{u_1\, u_2} } \right) 
\left( {\alpha}^4 -1 \right)    \right.      \notag \\
& \left.  +\left[ b_1 \left( u_1+ \frac{1}{u_1} \right) 
-b_2 \left( u_2+ \frac{1}{u_2} \right) \right] {\alpha}^2 \right\} ,    
\label{dFa} 
\end{align}
\begin{align} 
&  \frac{\partial F}{\partial u_1} =\frac{ {\alpha}^2 }{\left( {\alpha}^4 +1 \right) u_1 }  
\left[ b_1 \left( u_1 -\frac{1}{u_1} \right) {\alpha}^2    \right.   \notag \\
&  \left.  -\left( c\sqrt{u_1\, u_2} +\frac{d}{ \sqrt{u_1\, u_2} }\right) \right] ,
\label{dFu1} 
\end{align}
\begin{align} 
&  \frac{\partial F}{\partial u_2} =\frac{ {\alpha}^2 }{\left( {\alpha}^4 +1 \right) u_2 }  
\left[ b_2 \left( u_2 -\frac{1}{u_2} \right)  \frac{1}{ {\alpha}^2 }   \right.   \notag \\
&  \left.  -\left( c\sqrt{u_1\, u_2} +\frac{d}{ \sqrt{u_1\, u_2} }\right) \right]
\label{dFu2} 
\end{align}
vanish at a stationary point of the function\ (\ref{F3}). We try to solve 
the resulting system of stationarity equations:
\begin{align}  
&  \left( c\sqrt{u_1\, u_2} -\frac{d}{ \sqrt{u_1\, u_2} } \right) 
\left( 1-{\alpha}^4 \right)       \notag \\
& =\left[ b_1 \left( u_1+ \frac{1}{u_1} \right) 
-b_2 \left( u_2+ \frac{1}{u_2} \right) \right] {\alpha}^2 ,    
\label{sta} 
\end{align}
\begin{align} 
b_1 \left( u_1 -\frac{1}{u_1} \right) {\alpha}^2   
=c\sqrt{u_1\, u_2} +\frac{d}{ \sqrt{u_1\, u_2} },
\label{stu1} 
\end{align}
\begin{align} 
b_2 \left( u_2 -\frac{1}{u_2} \right)  \frac{1}{ {\alpha}^2 }  
=c\sqrt{u_1\, u_2} +\frac{d}{ \sqrt{u_1\, u_2} }.
\label{stu2} 
\end{align}
From Eqs.\ (\ref{stu1}) and\ (\ref{stu2}) it follows:
\begin{align}  
{\alpha}^4 =\frac{ b_2 \left( u_2 -\frac{1}{u_2} \right) }
{ b_1 \left( u_1 -\frac{1}{u_1} \right) }\, ;  
\label{alpha} 
\end{align}
\begin{align}   
& b_1 b_2  \left( u_1 -\frac{1}{u_1} \right) \left( u_2 -\frac{1}{u_2} \right)   \notag \\
& =\left( c\sqrt{u_1 u_2} +\frac{d}{ \sqrt{u_1 u_2} } \right)^2 \,;    
\label{u1u2} 
\end{align}
\begin{align}  
\left[ \Delta P_{+}(\alpha) \right]^2 =\left[ \Delta Q(\alpha) \right]^2.     
\label{eqvar} 
\end{align}
Insertion of Eqs.\ (\ref{stu1}) and\ (\ref{alpha}) into Eq.\ (\ref{sta}) 
gives the proportionality relation
\begin{align}
u_2={\gamma} u_1,    \qquad    \gamma:=\frac{b_2 c-b_1 d}{b_1 c-b_2 d} \leqq 1.
\label{gamma}
\end{align}
By substituting it into Eq.\ (\ref{u1u2}), we get a quadratic equation 
in the product $p:=u_1 u_2 \geqq 1:$
\begin{align}
& \left( b_1 b_2 -c^2 \right) p^2 -\left[ b_1 b_2\left( \gamma +\frac{1}{\gamma}\right)
+2cd  \right] p      \notag \\
& +\left( b_1 b_2 -d^2 \right) =0
\label{eqp}
\end{align}
with
\begin{align} 
\gamma +\frac{1}{\gamma}= 2+\frac{\left[ \left( b_1 -b_2 \right)  \left( c +d \right)  \right]^2 }
{\left( b_1 c-b_2 d \right) \left( b_2 c-b_1 d \right) }.
\label{sum}
\end{align}
Let us denote by ${\Delta}_p$ the discriminant of the quadratic trinomial 
in Eq.\ (\ref{eqp}) and let $p_{\pm}$ be its roots. Making use of Eq.\ (\ref{A5}), 
we find the relation 
\begin{align}
{\Delta}_p=\left[ \frac{b_1 b_2 \left( c^2 -d^2 \right) }
{\left( b_1 c -b_2 d \right) \left( b_2 c -b_1 d \right) } \right]^2 {\Delta}^{\rm PT}  \geqq 0.
\label{Dp}
\end{align}
Since $p_ {-} < 1$ for $c+d > 0$, the only acceptable solution of the quadratic 
equation\ (\ref{eqp}) is 
\begin{align}
p_{+}=\frac{c \left( b_1 b_2 -d^2 \right) -d \left( {\kappa}_{-}^{\rm PT} \right)^2 }
{-d \left( b_1 b_2 -c^2 \right) +c \left( {\kappa}_{-}^{\rm PT} \right)^2 }  \geqq 1.
\label{p+}
\end{align}
With Eqs.\ (\ref{gamma}) and\ (\ref{p+}) we get the scaling factors:
\begin{align}
& u_{1m} =\left[ \frac{b_2 \left( b_1 b_2 -d^2 \right) -b_1 \left( {\kappa}_{-}^{\rm PT} \right)^2 }
{b_2 \left(b_1 b_2 -c^2 \right) -b_1 \left( {\kappa}_{-}^{\rm PT} \right)^2 } \right]^{\frac{1}{2} },             
\notag    \\
& u_{2m} =\left[ \frac{b_1 \left( b_1 b_2 -d^2 \right) -b_2 \left( {\kappa}_{-}^{\rm PT} \right)^2 }
{b_1 \left(b_1 b_2 -c^2 \right) -b_2 \left( {\kappa}_{-}^{\rm PT} \right)^2 } \right]^{\frac{1}{2} }.
\label{u_jm}
\end{align}
In view of Eq.\ (\ref{A5}), they have the alternative expressions:
\begin{align}
& u_{1m} =\left\{ \frac{ b_1 \left[ \sqrt{ {\Delta}^{\rm PT} }-\left( b_1^2 -b_2^2 \right) \right]
+2d \left( b_1 c -b_2 d \right) }
{ b_1 \left[ \sqrt{ {\Delta}^{\rm PT} }-\left( b_1^2 -b_2^2 \right) \right]
-2c \left( b_2 c -b_1 d \right) } \right\}^{\frac{1}{2} },    
\notag    \\
& u_{2m} =\left\{ \frac{ b_2 \left[ \sqrt{ {\Delta}^{\rm PT} }+\left( b_1^2 -b_2^2 \right) \right]
+2d \left( b_2 c -b_1 d \right) }
{ b_2 \left[ \sqrt{ {\Delta}^{\rm PT} }+\left( b_1^2 -b_2^2 \right) \right]
-2c \left( b_1 c -b_2 d \right) } \right\}^{\frac{1}{2} }.
\label{u_km}
\end{align}
Insertion of  the scaling factors\ (\ref{u_km}) into  Eq.\ (\ref{alpha}) 
leads to the following value of the EPR-like parameter $\alpha$:
\begin{align}
{\alpha}_m=\left[ \frac{ \sqrt{ {\Delta}^{\rm PT} }-\left( b_1^2 -b_2^2 \right) }
{ \sqrt{ {\Delta}^{\rm PT} }+\left( b_1^2 -b_2^2 \right) } \right]^{\frac{1}{4} } \leqq 1.
\label{a_m}
\end{align}
There is therefore a single stationary value of the function 
$ F\left( {\alpha}^2,\, u_1,\, u_2 \right) $,  Eq.\ (\ref{F3}): 
\begin{align}
F_m = F\left( {\alpha}_m^2, u_{1m}, u_{2m} \right). 
\label{F_m}
\end{align}
In fact, this is its absolute minimum, as proven in Appendix B.
With the solution\ (\ref{u_jm}) and\ (\ref{a_m}), we take advantage 
of Eqs.\ (\ref{eqvar}), \ (\ref{A6}), and\ (\ref{A8}) to find the minimum value
\begin{align} 
F_m=2 {\kappa}_{-}^{\rm PT}. 
\label{Fmin}
\end{align}
This concludes the proof. Moreover, although stated for $d<0$, Eq.\ (\ref{Fmin}) 
is also true for $0 \leqq d < \frac{b_2}{b_1}c$.

It is worth mentioning that Eqs.\ (\ref{u_km}) and\ (\ref{a_m}), which specify 
the coordinates of the minimum point, become much simpler for two classes of TMGSs, 
namely, the STSs and the symmetric states. We find it convenient to write down here 
the corresponding formulae.
\begin{enumerate}

\item {\em Two-mode STSs}:
\begin{align}
& c=-d>0  \;\;  \Longleftrightarrow   \;\;  u_{1m}=u_{2m}=1:  \qquad   \gamma=1;    \notag \\
& \sqrt{ {\Delta}^{\rm PT} }=\left( b_1 +b_2 \right)  \sqrt{\delta},    \qquad    
\delta:=\left( b_1 -b_2 \right)^2 +4c^2 >0:         \notag \\
& \left( {\alpha}_m \right)^2= \frac{1}{2c}\left[  \sqrt{\delta}-\left( b_1 -b_2 \right) \right].
\label{STS}
\end{align}

\item {\em Symmetric TMGSs}:
\begin{align}
& b_1=b_2 =:b  \;\;  \Longrightarrow   \;\;  u_{1m}=u_{2m}=\sqrt{\frac{b+d}{b-c} }:  
\quad   \gamma=1;    \notag \\
& \sqrt{ {\Delta}^{\rm PT} }=2b(c-d) >0,    \qquad    {\alpha}_m =1.    
\label{sym}
\end{align}
\end{enumerate}
It is interesting to notice that in Ref.\cite{ASI} the minimal EPR-like uncertainty 
in sum form was explicitly evaluated for symmetric TMGSs. The
authors have interpreted therein the smallest symplectic eigenvalue of the partially 
transposed state as a quantifier of the greatest amount of EPR-like correlations 
which can be created in a symmetric TMGS by means of local operations. 
Our present findings in the general case, Eq.\ (\ref{Emin}) for the product form 
of the normalized EPR-like uncertainties, and Eq.\ (\ref{Fmin})
for the sum-form ones, appear to be consistent with this interpretation.
Therefore, the local squeezing factors\ (\ref{u_km}) determine the state 
having the maximal normalized  EPR-like correlations among the whole set of equally 
entangled TMGSs.

\subsection{Separability}

Let us write the necessary conditions of separability\ (\ref{EF}) for a TMGS 
by using the functions\ (\ref{E3}) and\ (\ref{F2}):
\begin{align}
E\left( \lambda,\, \mu  \right)  \geqq \frac{1}{4},   \qquad    
F\left( {\alpha}^2,\, u_1,\, u_2 \right) \geqq 1. 
\label{EFGS} 
\end{align}
When $d<0$, these inequalities hold for any values of the above functions,
including their minima\ (\ref{Emin}) and\ (\ref{Fmin}), respectively: 
\begin{align}
E_m=\left( {\kappa}_{-}^{\rm PT} \right)^2 \geqq \frac{1}{4},   \qquad    
F_m=2 {\kappa}_{-}^{\rm PT} \geqq 1. 
\label{EmFm} 
\end{align}
In the opposite case, $d \geqq 0$, the identity\ (\ref{d>0}) entails two equivalent 
inequalities:
\begin{align}
{\kappa}_{-}^{\rm PT} \geqq \frac{1}{2} \; \;  \Longleftrightarrow  \; \;  {\cal D}^{\rm PT} \geqq 0. 
\label{PS} 
\end{align} 
Therefore, each EPR-like necessary condition of separability\ (\ref{EFGS}) for a TMGS
implies the Peres-Simon inequality\ (\ref{DPT1}). We have reached this conclusion
making no use of the PPT theory.

On the other hand, Simon's necessary condition of separability\ (\ref{DPT1}) 
is also a sufficient one for any TMGS \cite{Simon}. When $d \geqq 0$, 
the inequalities\ (\ref{PS}) hold, so that the TMGS is separable.
The only debatable case is therefore $d<0$. We apply the PPT criterium
of separability\ (\ref{PS}) in conjunction with Eqs.\ (\ref{Emin}) and\ (\ref{Fmin}) 
to get two EPR-like conditions:
\begin{align}
E_m \geqq  \frac{1}{4},  \quad  F_m  \geqq 1 \;\;\Longleftrightarrow  \;\; 
{\hat \rho} \;\;  {\rm separable},   \quad    (d<0).      
\label{EFsep}
\end{align}
Equation\ (\ref{EFsep}) can be written alternatively:
\begin{align}
E_m < \frac{1}{4},  \quad  F_m < 1  \;\;  \Longleftrightarrow   \;\;   
{\hat \rho} \;\;  {\rm  entangled},     \quad    (d<0). 
\label{EFent}
\end{align}
Accordingly, both minima $E_m$ and $F_m$ are themselves
separability indicators. This means that each of the identities\ (\ref{EFGS})
is true if and only if the TMGS is separable.

\section{Separability indicator in regularized sum form}

\subsection{EPR-like correlation function}

Let us start from the Reid's pair of EPR-like observables\ (\ref{Reid}). One gets 
their variances by setting
${\alpha}_1=1,\; \\ {\alpha}_2=:\lambda,\; {\beta}_1=1,\;  {\beta}_2=:\mu$ 
in Eqs.\ (\ref{varQ}) and\ (\ref{varP}):
\begin{align}
\left( \Delta Q \right)^2=b_1 u_1 +b_2 u_2 {\lambda}^2 -2c \sqrt{u_1 u_2}\, \lambda,
\label{varQ1a} 
\end{align}
\begin{align}
\left( \Delta P \right)^2=\frac{b_1}{u_1} +\frac{b_2}{u_2}\,{\mu}^2 
+2\frac{d}{\sqrt{u_1 u_2} }\,\mu.
\label{varP1a} 
\end{align}
It is convenient to absorb the scaling factor $u_2$ into a pair 
of new EPR-like parameters:
\begin{align}
\xi:=\sqrt{u_2}\, \lambda >0,    \qquad    \eta:=\frac{1}{\sqrt{u_2} }\, \mu >0. 
\label{xi-eta} 
\end{align}
This simplifies the variances\ (\ref{varQ1a}) and\ (\ref{varP1a}) as follows:
\begin{align}
\left[ {\Delta Q}(\xi) \right]^2 =b_1 u_1 +b_2 {\xi}^2 -2c \sqrt{u_1}\, \xi,
\label{varQ3} 
\end{align}
\begin{align}
\left[ {\Delta P}(\eta) \right]^2 =\frac{b_1}{u_1} +b_2 {\eta}^2 
+2\frac{d}{\sqrt{u_1} }\,\eta.
\label{varP3} 
\end{align}
We take inspiration from the separability condition\ (\ref{sumsG}) to build 
an EPR-like correlation function of the remaining independent variables
$\xi,\, \eta,\, u_1:$
\begin{align}
G\left( \xi,\, \eta,\, u_1 \right):=\left[ {\Delta Q}(\xi) \right]^2 +\left[ {\Delta P}(\eta) \right]^2 
-\left( 1+\xi \eta \right).
\label{G} 
\end{align}
The function\ (\ref{G}), which is non-negative for any separable TMGS,
has the explicit form:                              
\begin{align}
& G\left( \xi,\, \eta,\, u_1 \right)=\left( b_1 u_1 +b_2 {\xi}^2 -2c \sqrt{u_1}\, \xi \right)     \notag \\
& +\left( \frac{b_1}{u_1} +b_2 {\eta}^2 +2\frac{d}{\sqrt{u_1} }\,\eta \right) -\left( 1+\xi \eta \right).
\label{G1} 
\end{align}
The following statement is true.

{\bf Theorem 3.} {\em For any TMGS with $d<0$ the function  
$G\left( \xi,\, \eta,\, u_1 \right)$ has an absolute minimum,
\begin{align}
G_m:=\min_{ \{ \xi,\, \eta,\, u_1 \} } G\left( \xi,\, \eta,\, u_1 \right),
\label{minG}
\end{align}
which is the product of the determinant ${\mathcal D}^{\rm PT}$
and a positive factor.}

{\em Proof}. Let us assume that $d<0$ in Eq.\ (\ref{G1}) and in the subsequent ones. 
We write the stationarity conditions:
\begin{align}   
& \frac{\partial G}{\partial \xi} =0:  \qquad    2b_2 \xi -\eta = 2c\sqrt{u_1},     \notag \\
& \frac{\partial G}{\partial \eta} =0:  \qquad   -\xi +2b_2 \eta =-2\frac{d}{\sqrt{u_1} }  \notag \\
&  \frac{\partial G}{\partial u_1} =0:  \qquad     c u_1 \xi +d \eta 
=\frac{b_1}{\sqrt{u_1} }\left( u_1^2 -1 \right).
\label{d1} 
\end{align}
The system\ (\ref{d1}) is linear in the variables\ (\ref{xi-eta}). By solving 
the first two equations with respect to $\xi$ and $\eta$ and then replacing 
the result into the third one, we find its unique solution:
\begin{align}
& {\xi}_m =\frac{1}{b_2^2 -\frac{1}{4} }\left(b_2 c \sqrt{u_{1m} }
-\frac{1}{2} \frac{d}{ \sqrt{u_{1m} } } \right),          \notag \\
& {\eta}_m =\frac{1}{b_2^2 -\frac{1}{4} } \left( \frac{1}{2} c \sqrt{u_{1m} } 
-b_2 \frac{d}{ \sqrt{u_{1m} } } \right),        \notag \\
& u_{1m}=\left[ \frac{b_2 \left( b_1 b_2-d^2 \right) -\frac{1}{4} b_1}
{b_2 \left( b_1 b_2-c^2 \right) -\frac{1}{4} b_1} \right]^{\frac{1}{2} }.
\label{mp}
\end{align}
In Appendix B we have proven that the only stationary value of the function\ (\ref{G1}),  
\begin{align}
G_m = G\left( {\xi}_m,\, {\eta}_{m},\, u_{1m} \right), 
\label{G_m}
\end{align}
is its absolute minimum. Insertion of the coordinates\ (\ref{mp}) into Eq.\ (\ref{G1}) 
and a subsequent straightforward calculation yields the formula
\begin{align}
& G_m=\frac{1}{b_2^2 -\frac{1}{4} }
\left( 2\left\{ \left[ b_2 \left( b_1 b_2-c^2 \right)-\frac{1}{4}b_1 \right]   \right. \right.  \notag \\
& \left. \left. \times \left[ b_2 \left( b_1 b_2-d^2 \right) -\frac{1}{4}b_1 \right] \right\}^{\frac{1}{2} }
-\left(b_2^2 -\frac{1}{4} -c d\right) \right).
\label{G_m1}
\end{align}
Making use of the identity\ (\ref{A10}), the minimum\ (\ref{G_m1}) can be cast into the form: 
\begin{align}
&G_m =4\,{\mathcal D}^{\rm PT} \left( 2\left\{ \left[ b_2 \left( b_1 b_2-c^2 \right)-\frac{1}{4}b_1 \right]   
\right. \right.  \notag \\
& \left. \left. \times \left[ b_2 \left( b_1 b_2-d^2 \right) -\frac{1}{4}b_1 \right] \right\}^{\frac{1}{2} }
+\left(b_2^2 -\frac{1}{4} -c d\right) \right)^{-1}.
\label{G_m2}
\end{align}
Because $d<0$, Eq.\ (\ref{G_m2}) displays the feature that the minimum value $G_m$ 
and the determinant ${\mathcal D}^{\rm PT}$ have the same sign. 
The proof is complete. 

\subsection{Separability}

We exploit the above-mentioned condition of separability:
when a given TMGS is separable, then the function\ (\ref{G1}) is non-negative:
\begin{equation}
G\left( \xi,\, \eta,\, u_1 \right) \geqq 0.
\label{G1>0}
\end{equation}
If $d<0$, the minimum\ (\ref{G_m2}) is therefore non-negative,
\begin{equation}
G_m \geqq 0,
\label{G_m>0}
\end{equation}
so that Simon's condition of separability\ (\ref{DPT}) is fulfilled.
As shown in Sec. IV, any state with $d \geqq 0$ observes the separability
condition\ (\ref{PS}). Put together, this means that the EPR-like necessary condition 
of separability\ (\ref{G1>0}) implies the Peres-Simon inequality\ (\ref{DPT1}). 
We have reached this conclusion without any use of the PPT idea.

Conversely, according to Simon's separability criterion\ (\ref{DPT1}) any TMGS 
with $d \geqq 0$ is separable. By virtue of Eq.\ (\ref{G_m2}), it also implies that
a TMGS with $d<0$ is separable if the inequality\ (\ref{G_m>0}) is satisfied.
As this condition is already proven to be necessary, it follows that the minimum
$G_m$ is a separability indicator in its own right:
\begin{align}
G_m \geqq 0 \;\;\Longleftrightarrow  \;\; {\hat \rho} \;\;  {\rm separable}.
\label{Gsep}
\end{align}
Equation\ (\ref{Gsep}) has the obvious alternative form:
\begin{align}    
& G_m <0  \;\;  \Longleftrightarrow   \;\;   {\hat \rho} \;\;  {\rm  entangled}.     
\label{Gent}
\end{align}
Finally, on account of Theorem 3, Simon's separability criterion for a TMGS, 
Eq.\ (\ref{PS}), entails that the identity\ (\ref{G1>0}) is not only a necessary, 
but also a sufficient condition of separability.

\section{The EPR-like approach to separability revisited}  

At this point it is useful to re-examine the full EPR-like approach to the separability
of TMGSs put forward by Duan {\em et al.} in Ref. \cite{Duan}. 

\subsection{Basic equations} 

We start from the variances of the non-local one-parameter observables\ (\ref{Duan}) 
given by Eqs.\ (\ref{varQ4}) and\ (\ref{varP4}). Let us introduce the non-local operator
\begin{align}
& {\hat P}(\alpha):=
\begin{cases}
{\hat P}_{+}(\alpha),  \qquad    (d<0),  \\
{\hat P}_{-}(\alpha),   \qquad    (d \geqq 0).  \\
\end{cases}
\label{hatP} 
\end{align}
Its variance,
\begin{align}
\left[ \Delta P(\alpha) \right]^2 =\frac{b_1}{u_1} {\alpha}^2 
+\frac{b_2}{u_2} \frac{1}{ {\alpha}^2 }-\frac{2|d|}{ \sqrt{u_1\, u_2} },
\label{varPm} 
\end{align}
is the minimum over the pair of variances\ (\ref{varP4}) with respect to the sign of $d$:
\begin{align} 
\Delta P(\alpha)=\min_{ \{ {\rm sgn}(d) \} }\{ \Delta P_{+}(\alpha), \, \Delta P_{-}(\alpha) \}.
\label{varPmin} 
\end{align}
Recall the definition of the signum function of a real variable:
\begin{align}
& {\rm sgn}(x):=
\begin{cases}
-1,  \qquad    (x<0),  \\
\;\;\; 0,  \qquad    (x=0),  \\
\;\;\; 1,  \qquad    (x>0).   \notag
\end{cases}
\end{align}
Guided by the necessary conditions of separability\ (\ref{sumsD}) derived 
by Duan {\em et al.} in Ref. \cite{Duan}, we employ the following 
EPR-like correlation function, which is a regularized sum depending on the local scalings $u_1, u_2$:
\begin{align}
& K\left( {\alpha}^2,\, u_1,\, u_2 \right) := \left[ \Delta Q(\alpha) \right]^2 
+\left[ \Delta P(\alpha) \right]^2             \notag \\ 
& -\left( {\alpha}^2 +\frac{1}{{\alpha}^2} \right).
\label{K} 
\end{align}
Substitution of the variances\ (\ref{varQ}) and\ (\ref{varP}) into Eq.\ (\ref{K}) 
gives the explicit formula:
\begin{align} 
& K\left( {\alpha}^2,\, u_1,\, u_2 \right) 
={\alpha}^2 \left[ b_1 \left( u_1+ \frac{1}{u_1} \right) -1 \right]       \notag \\
& +\frac{1}{ {\alpha}^2 } \left[ b_2 \left( u_2+ \frac{1}{u_2} \right) -1 \right]       \notag \\
& -2\left( c\sqrt{u_1\, u_2} +\frac{|d|}{ \sqrt{u_1\, u_2} }\right).
\label{K1} 
\end{align}

We  apply the same pattern as in Sec. V, 
starting from the necessary condition of separability\ (\ref{sumsD}) which reads:
\begin{align}
K\left( {\alpha}^2,\, u_1,\, u_2 \right) \geqq 0.
\label{K>0} 
\end{align}
Our approach is to minimize the function\ (\ref{K1}). This reaches its minimum 
with respect to the variable ${\alpha}^2$ for the value
\begin{align}  
{\tilde \alpha}_m^2 =\sqrt{ \frac{ b_2 \left( u_2 +\frac{1}{u_2} \right) -1}
{ b_1 \left( u_1 +\frac{1}{u_1} \right) -1} }\, .  
\label{alpha_m} 
\end{align}
The obtained minimum is a function of the scaling factors:
\begin{align} 
& f\left( u_1,\, u_2 \right) :=K\left( {\tilde \alpha}_m^2,\, u_1,\, u_2 \right)          \notag \\
& =2\left\{ \left[ b_1 \left( u_1+ \frac{1}{u_1} \right) -1 \right]    
\left[ b_2 \left( u_2+ \frac{1}{u_2} \right) -1 \right] \right\}^{ \frac{1}{2} }     \notag \\
& -2\left( c\sqrt{u_1\, u_2} +\frac{|d|}{ \sqrt{u_1\, u_2} }\right).
\label{f(uu)} 
\end{align}
The stationarity conditions for the function\ (\ref{f(uu)}) reduce to the following 
system of equations in the unknowns $u_1$ and $u_2$:
\begin{align} 
\frac{ \frac{b_1}{u_1}- \frac{1}{2} }{ b_1 u_1 -\frac{1}{2} }
=\frac{ \frac{b_2}{u_2}- \frac{1}{2} }{ b_2 u_2 -\frac{1}{2} },      
\label{eq1} 
\end{align}
\begin{align} 
& \sqrt{\left( b_1 u_1 -\frac{1}{2} \right) \left( b_2 u_2 -\frac{1}{2} \right) }
-\sqrt{\left( \frac{b_1}{u_1}-\frac{1}{2} \right) \left( \frac{b_2}{u_2}-\frac{1}{2} \right) }
\notag \\
& = c \sqrt{ u_1 u_2}  -\frac{ |d| }{ \sqrt{ u_1 u_2} }. 
\label{eq2} 
\end{align}
Making use of Eq.\ (\ref{eq1}),  it is convenient to replace Eq.\ (\ref{eq2})
by a polynomial one:
\begin{align} 
b_1 b_2 \left( u_1^2 -1 \right)  \left( u_2^2 -1 \right) =\left( c u_1 u_2 -|d| \right)^2 .
\label{eq3} 
\end{align}
Equations\ (\ref{eq1}) and\ (\ref{eq2}) coincide with those written by Duan {\em et al.} 
in Ref. \cite{Duan} in order to define what they have called the standard form II 
of the CM of a TMGS. We prefer instead the name {\em witness
standard form of the CM.As it will be shown in Sec. VI C, this} 
seems to be more appropriate.  We also term the local squeezings 
$\left\{ {\tilde u}_1, {\tilde u}_2 \right\}$ as {\em witness scaling factors}. 

However, when trying to solve analytically the system under discussion 
in the general case, one faces a non-trivial algebraic equation of degree eight. 
For the time being, one does not know an explicit solution  
$\left\{ {\tilde u}_1,\, {\tilde u}_2 \right \}$ of Eqs.\ (\ref{eq1}) and\ (\ref{eq3}), 
except for some special classes of TMGSs, such as the thermal states (TSs), 
the symmetric ones \cite{PT2008}, the mode-mixed thermal states (MTSs), 
and the STSs \cite{PT2015,PT2016}, as well as the states subject
to the constraint ${\mathcal D}:=\det \left( {\mathcal V}+\frac{i}{2}J \right) =0$ \cite{ASI}. 
We also remark that the scaling factors  written in Refs.\cite{K2009,K2012} 
are valid only for TMGSs at the separability threshold, i. e., fulfilling the condition 
${\mathcal D}^{\rm PT}=0$.
\\
\subsection{Existence of a witness standard form 
${\mathcal V}\left( {\tilde u}_1, {\tilde u}_2 \right)$ of the covariance matrix} 

It is extremely helpful to state here a theorem that sharpens a result of Ref. \cite{Duan} 
regarding the existence of a solution of the above system in the general case. 

{\bf Theorem 4.} {\em For any TMGS  there exists at least a solution 
$\left\{ {\tilde u}_1,\, {\tilde u}_2 \right\}$ of the algebraic system\ (\ref{eq1}) 
and\ (\ref{eq3}) in the classicality range of the local squeeze factors: 
$${\tilde u}_1 \in [1, 2b_1], \quad  {\tilde u}_2 \in [1, 2b_2]. $$
Such a solution determines awitness} standard form of the CM:  
${\mathcal V}\left( {\tilde u}_1, {\tilde u}_2 \right)$.

{\em Proof}. Note that Eq.\ (\ref{eq1}) displays the pairings:
\begin{align}
u_1= 1  \;  \Longleftrightarrow  \;  u_2= 1,  \qquad
u_1= 2b_1  \;  \Longleftrightarrow  \;  u_2= 2b_2.    
\label{pairs} 
\end{align}
As a matter of fact, Eq.\ (\ref{eq1}) is a quadratic one in each of the variables 
$u_1$ and $u_2$, leading to a bijective continuous function,
\begin{align}
u_2=h(u_1),  \qquad  h:[1, 2b_1]  \;  \longrightarrow  \; [1, 2b_2],
\label{h} 
\end{align}
which reads:
\begin{align}
& h(u_1)=\frac{ 2b_{2}u_{1}\left( 2b_{1}u_{1}-1 \right) }
{ b_{1}\left( u_{1}^2 -1 \right) +\sqrt{{\Delta}_1} },     \notag \\
& {\Delta}_1:= b_{1}^2\left( u_{1}^2 -1 \right)^2+4 b_{2}^2 u_{1}
\left( 2b_1- u_1 \right) \left( 2b_{1} u_{1}-1 \right).
\label{h(u_1)} 
\end{align}
One gets the inverse function
\begin{align}
u_1=h^{-1}(u_2),  \qquad  h^{-1}:[1, 2b_2]  \;  \longrightarrow  \; [1, 2b_1],
\label{h^{-1}} 
\end{align}
by interchanging the mode indices 1 and 2  in Eq.\ (\ref{h(u_1)}).
The bijective function $h(u_1)$, Eq.\ (\ref{h(u_1)}), is strictly and continuously 
increasing from the initial value $h(1)=1$ to the final one $h(2b_1)=2b_2$.
Remark that for a symmetric TMGS it reduces to the identity function: 
\begin{align}
u_2= u_1,  \qquad  (b_1=b_2=:b).
\label{h=1} 
\end{align}
Coming back to the general case, we introduce a function of two variables 
suggested by Eq.\ (\ref{eq3}):
\begin{align} 
\Phi(u_1, u_2):= b_1 b_2 \left( u_1^2 -1 \right)  \left( u_2^2 -1 \right) 
-\left( c u_1 u_2 -|d| \right)^2 .
\label{Phi} 
\end{align}
The existence of a solution of the algebraic system\ (\ref{eq1}) and\ (\ref{eq3}) 
is equivalent to that of a zero of the one-variable function
 \begin{align} 
\phi(u_1):=\Phi\left( u_1, h(u_1) \right),  \qquad    (\, u_1 \in [1, 2b_1] \,).
\label{phi} 
\end{align} 
In order to check if such a zero exists indeed, we have to examine the sign 
of the values of the function\ (\ref{phi}) at the end points of its domain:
\begin{align} 
\phi(1)=\Phi(1,1)=-(c-|d|)^2:   \qquad   \phi(1) \leqq 0;
\label{phi1} 
\end{align}
\begin{align} 
& \phi(2b_1)=\Phi\left( 2b_1, 2b_2 \right)\nonumber\\=&16b_1 b_2 \left[ {\mathcal D}
+d^2 \left( b_1 b_2-c^2-\frac{1}{4} \right) +\frac{1}{2}c(|d|+d) \right]     \notag  \\
& +4d^2\left( b_1 b_2 -\frac{1}{4} \right):   \qquad   \phi(2b_1) \geqq 0.
\label{phi2} 
\end{align}
The inequality in Eq.\ (\ref{phi2}) stems from conditions\ (\ref{GS1})-\ (\ref{GS3}),
which are equivalent to the Robertson-Schr\"odinger uncertainty relation, Eq.\ (\ref{RS}).
Therefore, by virtue of continuity, the function\ (\ref{phi}) has at least a zero 
in the classicality interval  $u_1 \in [1, 2b_1]$ for any TMGS. We denote
the corresponding solution of Eqs.\ (\ref{eq1}) and\ (\ref{eq3}) 
by $\{ \tilde{u}_1, \tilde{u}_2=h(\tilde{u}_1) \}$, so that 
\begin{align} 
& \phi( \tilde{u}_1)=\Phi( \tilde{u}_1, \tilde{u}_2)=0,         \notag  \\  
& (\, \tilde{u}_1 \in [1, 2b_1], \;  \tilde{u}_2 \in [1, 2b_2]\,).
\label{phi3} 
\end{align}
This concludes the proof.

\subsection{Separability indicator} 

The witness standard form of the CM is important because the condition 
of classicality for the corresponding TMGS,
\begin{equation}
{\mathcal V}\left( {\tilde u}_1, {\tilde u}_2 \right)-\frac{1}{2}I_4 \geqslant 0,
\label{class}
\end{equation}
is equivalent to the requirement
\begin{equation}
{\tilde f}:=f\left( {\tilde u}_1, {\tilde u}_2 \right) \geqq 0.
\label{f>0}
\end{equation}
Indeed, taking account of Eqs.\ (\ref{eq1}) and\ (\ref{eq3}), we get the twin formulae:
\begin{align} 
& {\tilde f}=4\left[ \sqrt{\left( b_1{\tilde u}_1 -\frac{1}{2} \right) 
\left( b_2{\tilde u}_2 -\frac{1}{2} \right) } -c\sqrt{ {\tilde u}_1{\tilde u}_2 } \right],     \notag \\
& {\tilde f}=4\left[ \sqrt{ \left( \frac{b_1}{ {\tilde u}_1} -\frac{1}{2} \right) 
\left( \frac{b_2}{ {\tilde u}_2 } -\frac{1}{2} \right) }
-\frac{ |d| }{ \sqrt{ {\tilde u}_1{\tilde u}_2 } } \right].
\label{twin} 
\end{align}
We find it suitable to introduce the parallel notations:
\begin{align} 
& {\tilde f}^{\prime}:=4\left[ \sqrt{\left( b_1{\tilde u}_1 -\frac{1}{2} \right) 
\left( b_2{\tilde u}_2 -\frac{1}{2} \right) } +c\sqrt{ {\tilde u}_1{\tilde u}_2 } \right] >0,    \notag \\
& {\tilde f}^{\prime \prime}:=4\left[ \sqrt{ \left( \frac{b_1}{ {\tilde u}_1} -\frac{1}{2} \right) 
\left( \frac{b_2}{ {\tilde u}_2 } -\frac{1}{2} \right) }
+\frac{ |d| }{ \sqrt{ {\tilde u}_1{\tilde u}_2 } } \right] \geqq 0.
\label{twin1} 
\end{align}
The matrix condition\ (\ref{class}) reduces to four inequalities: 
\begin{align}  
& b_1{\tilde u}_1 -\frac{1}{2} \geqq 0,       \notag \\
& \left( b_1{\tilde u}_1 -\frac{1}{2} \right) \left( \frac{b_1}{ {\tilde u}_1 } -\frac{1}{2} \right) 
\geqq 0,            \notag \\
& \left( \frac{b_1}{ {\tilde u}_1 } 
-\frac{1}{2} \right) 4^{-2}({\tilde f} {\tilde f}^{\prime}) \geqq 0,       \notag \\
& \det\left[ {\mathcal V}\left( {\tilde u}_1, {\tilde u}_2 \right)-\frac{1}{2}I_4 \right]
=4^{-4} ({\tilde f} {\tilde f}^{\prime} )({\tilde f} {\tilde f}^{\prime \prime} ) \geqq 0.
\label{ineq} 
\end{align}
Three of them are already satisfied, so that the only condition to be fulfilled 
remains ${\tilde f} \geqq 0$, Eq.\ (\ref{f>0}). This classicality requirement 
is a sufficient condition for the separability of the given TMGS ${\hat \rho}$, 
whose CM is congruent with its witness standard form 
${\mathcal V}\left( {\tilde u}_1, {\tilde u}_2 \right) $ via a local symplectic 
transformation, Eq.\ (\ref{local}).  Moreover, by virtue of Eq.\ (\ref{K>0}), 
the inequality\ (\ref{f>0})  is also a necessary condition of separability.
Therefore, a TMGS whose CM has the {\em witness
standard form}  is a unique state 
for which separability reduces to classicality. By examining the sign 
of the EPR-like correlation function ${\tilde f}:=f\left( {\tilde u}_1, {\tilde u}_2 \right)$, 
one can check whether the witness-standard-form TMGS is classical or not, i.e., 
whether it possesses or not a well-behaved Glauber-Sudarshan $P$ representation.  

To sum up, our optimization method has exploited the EPR-like correlation 
function\ (\ref{K}) leading to the separability indicator ${\tilde f}$, Eq.\ (\ref{twin}). 
It is worth stressing that this indicator differs in two respects from the previous ones 
which are specified by Eqs.\ (\ref{Emin}),\ (\ref{Fmin}), and \ (\ref{G_m2}). 
First, it has been identified independently of Simon's PPT separability
criterion. Second, by contrast to the three indicators quoted above, 
we have not found yet in generality an explicit formula for the existing 
stationary point. We also are not quite sure that this is unique or that it corresponds 
to a minimum.

\subsection{Explicit witness scaling factors for special classes of states} 

It is instructive to present concisely the special classes of TMGSs whose witness 
scaling factors can be explicitly evaluated, as mentioned in Sec. VI A.

\begin{enumerate}

\item {\em TMGSs with} $c=|d| $: TSs, MTSs, and STSs.  \\ 
The solution 
\begin{align}
{\tilde u}_1 = {\tilde u}_2 =1
\label{DuanSTS}
\end{align}
of Eqs.\ (\ref{eq1}) and\ (\ref{eq3}) is specific to this class, which consists 
of TSs $(c=d=0)$, MTSs $(c=d > 0),$ and STSs $(c=-d > 0)$.  We have proven 
the uniqueness of the solution\ (\ref{DuanSTS}). Note the minimum value\ (\ref{alpha_m}): 
$${\tilde \alpha}_m^2 =\sqrt{ \frac{ b_2 -\frac{1}{2} }{ b_1 -\frac{1}{2} } }.$$
Equations\ (\ref{twin}) and\ (\ref{twin1}) give the separability indicator
\begin{align} 
& {\tilde f}=4\left[ \sqrt{\left( b_1-\frac{1}{2} \right) \left( b_2 -\frac{1}{2} \right) } -c \right],
\label{fSTS} 
\end{align}
and, respectively, the functions
\begin{align} 
& {\tilde f}^{\prime}={\tilde f}^{\prime\prime}=4\left[ \sqrt{\left( b_1-\frac{1}{2} \right) 
\left( b_2 -\frac{1}{2} \right) } +c \right].
\label{twin2} 
\end{align}
Equation\ (\ref{fSTS}) becomes insightful when written in terms of the symplectic 
eigenvalues ${\kappa}_{\pm}$ for an MTS and  ${\kappa}_{\pm}^{\rm PT}$ for an STS: 
 \begin{align} 
{\tilde f}_{\rm MT}=\frac{4^2}{ {\tilde f}^{\prime} } \left( {\kappa}_{-}
-\frac{1}{2}\right) \left( {\kappa}_{+}-\frac{1}{2}\right) \geqq 0,   \quad   (d>0),
\label{fMT1} 
\end{align}
\begin{align} 
{\tilde f}_{\rm ST}=\frac{4^2}{ {\tilde f}^{\prime} }\left( {\kappa}_{-}^{\rm PT}
-\frac{1}{2}\right) \left( {\kappa}_{+}^{\rm PT}-\frac{1}{2} \right),  \quad   (d<0).
\label{fST1} 
\end{align}

\item {\em Symmetric TMGSs}:   $\, b_1=b_2=:b.  \\$
We get a unique  solution,
\begin{align}
{\tilde u}_1 ={\tilde u}_2 =\sqrt{ \frac{b-|d|}{b-c} },
\label{Duansym}
\end{align}
which is specific to symmetric TMGSs, provided that $c > |d|.$ The minimum 
value\ (\ref{alpha_m}) of the parameter $\alpha$ is ${\tilde \alpha}_m =1$.  
We have proven the property of the solution\ (\ref{Duansym}) of being a minimum
 point of the function 
 (\ref{f(uu)}) 
written with $\, b_1=b_2=:b$. 
The corresponding absolute minimum is the EPR-like separability indicator
\begin{align} 
{\tilde f}=4\left[ \sqrt{(b-c)(b-|d|)}-\frac{1}{2} \right].
\label{fsym} 
\end{align}
According to Eqs.\ (\ref{kappa+-}) and\ (\ref{A5}),  
\begin{align} 
{\kappa}_{-}= \sqrt{(b-c)(b-d)},  \qquad    {\kappa}_{-}^{\rm PT}= \sqrt{(b-c)(b+d)},
\label{kappa-} 
\end{align}
so that the following relations hold:
\begin{align} 
{\tilde f}=4\left( {\kappa}_{-} -\frac{1}{2} \right) \geqq 0,     \qquad   (d \geqq 0),
\label{fsym1}  
\end{align}
\begin{align} 
{\tilde f}=4\left( {\kappa}_{-}^{\rm PT} -\frac{1}{2} \right),        \qquad  (d<0).
\label{fsym2} 
\end{align} 

\item {\em TMGSs at the boundary ${\tilde f}=0.$} \\
It will be shown in the next section that the property 
\begin{align}
{\tilde f}= 0    
\label{f=01}
\end{align}
is specific to all TMGSs with $d \geqq 0$ that are at the physicality edge 
$({\mathcal D}=0)$ and to those with $d<0$ that are at the separability threshold 
$\left( {\mathcal D}^{\rm PT}=0 \right).$ For both limit situations, we find 
a unique solution of the algebraic system\ (\ref{eq1}) and\ (\ref{eq3}):
\begin{align}
& {\tilde u}_1=2\frac{ c\left( b_{1}b_{2}-d^2 \right) +\frac{1}{4}|d| }{b_{1}|d|+b_{2}c},    \notag \\
& {\tilde u}_2 =2\frac{ c\left( b_{1}b_{2}-d^2 \right) +\frac{1}{4}|d| }{b_{1}c+b_{2}|d|}. 
\label{f=0}
\end{align} 
\end{enumerate}

Let us focus on the expressions\ (\ref{fMT1}),\ (\ref{fST1}), 
and\ (\ref{fsym1})-\ (\ref{fsym2}) of the EPR-like correlation function ${\tilde f}$
for MTSs, STSs, and symmetric TMGSs, respectively. These important special GSs
illustrate both cases: $d \geqq 0$ and $d<0$. The above explicit formulae clearly 
display the equivalence between the EPR-like separability condition\ (\ref{f>0}) 
and the PPT one, Eq.\ (\ref{PS}). However, such explicit expressions are not available
for an arbitrary TMGS. 

\section{Conditions of separability: EPR-like versus PPT}

In principle, the EPR-like separability criterion\ (\ref{f>0}) is as important 
as Simon's condition of separability\ (\ref{PS}) that relies on partial transposition.
Nevertheless, it suffers from the drawback that, in the general case, it cannot be 
handled analytically. This makes Simon's PPT criterion of separability\ (\ref{PS}) 
to prevail in practice. However, the next theorem explicitly connects 
the EPR-like inequality\ (\ref{f>0}) to the PPT one\ (\ref{PS}). 

{\bf Theorem 5.} {\em The EPR-like separability condition\ (\ref{f>0}) 
and Simon's PPT one, Eq.\ (\ref{PS}), are fully and manifestly equivalent.}

{\em Proof}. We introduce the following function, which is symmetric 
in the mode indices 1 and 2:
\begin{align}
& Z({u}_1, {u}_2):=\frac{1}{2}\left\{ \left[ \left( b_1{u}_1 -\frac{1}{2} \right) 
\left( b_2{u}_2 -\frac{1}{2} \right) - c^2 { u}_1 u_2  \right]  \right.     \notag \\ 
&  \times \left.   \left[\left( \frac{b_1}{ u_1} +\frac{1}{2} \right) \left(\frac{ b_2}{u_2} 
+\frac{1}{2} \right) - \frac{d^2 }{ u_1 u_2 } \right]  \right.     \notag \\ 
& +\left. \left[ \left( b_1 u_1 +\frac{1}{2} \right) 
\left( b_2{u}_2 +\frac{1}{2} \right) - c^2 { u}_1 u_2  \right]   \right.     \notag \\    
& \times \left. \left [\left( \frac{b_1}{ u_1} -\frac{1}{2} \right) \left(\frac{ b_2}{u_2} 
-\frac{1}{2} \right) - \frac{d^2 }{ u_1 u_2 } \right] \right\}. 
\label{Z}
\end{align}
The expression on the r. h. s. of Eq.\ (\ref{Z}) can be cast into a simpler form:
\begin{align}
Z({u}_1, {u}_2)=H(d){\mathcal D}+H(-d){\mathcal D}^{\rm PT} 
+\frac{1}{ 4u_1 u_2 } \Phi(u_1, u_2),
\label{Z1}
\end{align}
where $H(x):=\frac{1}{2}[1+{\rm sgn}(x)]$ denotes the Heaviside step function 
and $\Phi(u_1, u_2)$ is the function\ (\ref{Phi}). In view of  Eqs.\ (\ref{GS3}) 
and\ (\ref{DPT}), we write the formula
\begin{align}
H(d){\mathcal D}+H(-d){\mathcal D}^{\rm PT}=\left( b_1 b_2-c^2 \right) \left( b_1 b_2-d^2 \right)
\nonumber\\    
-\frac{1}{4} \left( b_1^2 +b_2^2+2c|d| \right) +\frac{1}{16}.  
\label{DD}
\end{align} 
Let us equate the expressions\ (\ref{Z}) and\ (\ref{Z1}) of the value 
$Z({\tilde u}_1, {\tilde u}_2)$ corresponding to a witness standard form  
${\mathcal V}\left( {\tilde u}_1, {\tilde u}_2 \right)$ of the CM. Taking account 
of Eqs.\ (\ref{twin}),\ (\ref{twin1}), and\ (\ref{phi3}), we get the identity:
\begin{align}
& H(d){\mathcal D}+H(-d){\mathcal D}^{\rm PT}    \notag \\
& =\frac{1}{32}{\tilde f} \left\{ {\tilde f}^{\prime}
\left[ \left( \frac{b_1}{\tilde u_1} +\frac{1}{2} \right) 
\left( \frac{ b_2}{\tilde u}_2 +\frac{1}{2} \right) 
- \frac{d^2 }{ {\tilde u}_1{\tilde u}_2 } \right]  \right.  \notag \\  
& \left. +{\tilde f}^{\prime\prime} \left[\left( b_1 \tilde u_1 +\frac{1}{2} \right) 
\left( b_2{\tilde u}_2 +\frac{1}{2} \right) - c^2 {\tilde u}_1{\tilde u}_2 \right] \right\}.
\label{id2}
\end{align}
In Eq.\ (\ref{id2}), the expression in curly brackets is strictly positive and  
${\mathcal D} \geqq 0$ for any TMGS ${\hat \rho}$.  Accordingly, the separability indicator 
${\tilde f}$ and the local symplectic invariant\ (\ref{DD}) do have the same sign. 
There are two cases:
\begin{itemize}
\item $d \geqq 0. \;\; {\rm Then} \;\; {\tilde f}\geqq 0 \;\; {\rm and} \;\;
{\mathcal D}^{\rm PT}\geqq 0:  \;\;  {\hat \rho}$  \; separable;
\item $d < 0.$  
\begin{align}
& {\rm Either} \;\; {\tilde f} \geqq 0\; \Longleftrightarrow \; 
    {\mathcal D}^{\rm PT} \geqq 0:  \quad{\hat \rho} \;   \; {\rm separable},   \notag \\ 
& {\rm or} \;\; {\tilde f}<0 \; \Longleftrightarrow \; {\mathcal D}^{\rm PT} < 0: 
\quad {\hat \rho}\;  \;\; {\rm entangled}.    \notag 
\end{align}
\end{itemize}
As the signs of the separability indicators $\tilde f$ and ${\mathcal D}^{\rm PT}$ 
coincide, we conclude they are equivalent in detecting separability of TMGSs:
\begin{align}
& {\tilde f} \geqq 0 \;\;  \Longleftrightarrow  \;\; {\mathcal D}^{\rm PT}  \geqq 0:    
    \quad {\hat \rho}\;  \;\; {\rm separable};     \notag \\ 
& {\tilde f} <0  \;\;  \Longleftrightarrow  \;\; {\mathcal D}^{\rm PT} < 0:   
\quad {\hat \rho}\; \;\; {\rm entangled}.
\label{fGm}
\end{align}
The proof is complete. Note the equivalence
\begin{align}
{\tilde f}= 0 \;\;  \Longleftrightarrow  \;\; H(d){\mathcal D}+H(-d){\mathcal D}^{\rm PT}=0,   
\label{fDD}
\end{align}
showing that the boundary  ${\tilde f}= 0$ consists of all TMGSs with $d \geqq 0$
that are at the physicality edge $({\mathcal D}=0)$ as well as of all those with $d<0$
that are at the separability threshold $\left( {\mathcal D}^{\rm PT}=0 \right).$ 

Theorem 5 enforces the idea that the two approaches 
to the separability problem for TMGSs analyzed  in this paper are fully equivalent. 
In spite of some recent  assertions \cite{K2009,K2012} regarding the weakness 
of the EPR-like treatment in comparison to the PPT one, we have succeeded
to prove their equivalence even in the absence of an explicit general solution 
for the witness scaling factors $\{ \tilde u_1, \tilde u_2 \}$. We also emphasize 
that Theorem 5 enables us to derive alternatively Simon's condition of separability 
for TMGSs without making any reference to the PPT formalism. 
Anyway, in order to decide whether a given TMGS is separable or not, then, 
by virtue of Eq.\ (\ref{fGm}), one is entitled to employ Simon's condition 
of separability, ${\mathcal D}^{\rm PT} \geqq 0,$ instead of the less efficient formula 
${\tilde f} \geqq 0$.

\section{Summary and conclusions} 

This work is devoted to an explicit application of EPR-like correlations 
in detecting Gaussian entanglement.  First, we tackle three correlation functions 
built with variances of two EPR-like observables in a TMGS. They are 
a normalized product,  a normalized sum, and a non-normalized, but regularized sum. 
The corresponding analytic results are Theorems 1, 2, and 3, which express 
their absolute minima in terms of either the smallest symplectic eigenvalue 
${\kappa}_{-}^{\rm PT}$ or the Simon determinant  ${\mathcal D}^{\rm PT}$. 
These lower bounds are explicitly written in Eqs.\ (\ref{Emin}), \ (\ref{Fmin}), 
and\ (\ref{G_m2}). On the one hand, we exploit them to point out that each 
of the three distinct EPR-like necessary conditions of separability for a TMGS 
implies the Peres-Simon PPT condition, Eq.\ (\ref{PS}). However, the EPR-like 
conditions are weaker than the latter because they do not imply its sufficiency. 
On the other hand, all three EPR-like correlations are separability indicators 
on account of Simon's PPT criterion of separability.

Due to their explicit expressions, Eqs.\ (\ref{Emin}) and \ (\ref{Fmin}), both minimal 
normalized EPR-like uncertainties, besides being separability markers, 
give a definite physical significance to the smallest symplectic eigenvalue 
${\kappa}_{-}^{\rm PT}$ of the CM of the partially transposed density matrix 
${\hat \rho}^{\rm PT}$. Indeed, ${\kappa}_{-}^{\rm PT}$ appears to be 
a quantifier of the greatest amount of EPR-like correlations that can be created
in any TMGS by means of local operations. 

Second,  we analyze the EPR-like approach of Duan {\em et al.} \cite{Duan} 
by applying an optimization method which is similar to those employed 
in Secs. IV and V. This leads us to a new separability indicator for TMGSs. 
We then prove straightforwardly that the corresponding necessary 
and sufficient EPR-like condition of separability of a TMGS is fully equivalent 
to Simon's PPT one \cite{Simon}.

The EPR-like correlation function introduced in Ref. \cite{Duan} is used in Sec. VI 
in a regularized sum form. An important conclusion is that its optimal value 
over the variables $\alpha, u_1,u_2,$ denoted ${\tilde f}$, turns out to be a separability 
indicator for TMGSs.  Recall that, among the original ideas put forward 
in Ref. \cite{Duan}, the central one is the existence of a witness standard form 
of the CM, which is confirmed here via Theorem 4. The requirement of classicality 
for the corresponding privileged TMGS is equivalent to the separability of the whole set 
of TMGSs connected to it by local unitary transformations. Thus, the separability 
properties of the whole class of TMGSs having a given set of standard-form parameters 
are assigned to this witness-standard-form state. Our EPR-like separability indicator 
${\tilde f}$, Eq.\ (\ref{twin}), is therefore obtained as a marker of classicality 
for the witness-standard-form TMGS. The resulting separability criterion\ (\ref{f>0}) 
is quite special: indeed, it is independent of the PPT condition, by contrast to  
the preceding three separability criteria\ (\ref{EFsep}) and\ (\ref{Gsep}). However, 
in spite of its soundness, the original EPR-like approach \cite{Duan} cannot decide 
analytically whether a TMGS is separable or not, except for the special cases discussed 
in Subsec. VI D. This happens because it does not provide a general analytic solution. 

Another main result is Theorem 5, which explicitly proves that the EPR-like indicator 
of separability ${\tilde f}$ is equivalent to Simon's PPT separability marker 
${\mathcal D}^{\rm PT}$. We have found a formula, Eq.\ (\ref{id2}), which reveals 
a direct connection between two distinct approaches to the separability problem 
for TMGSs, providing valuable insight onto it. At the same time, it constitutes
an alternative proof of Simon's PPT separability condition via the EPR-like
correlation-function method. This new perspective might stimulate further research
concerning the central role of the uncertainty relations in quantum mechanics.
Needless to say, the importance of the EPR-like approach is enhanced 
by our simple proof of its manifest consistency with Simon's PPT separability condition, 
whose practical usefulness is universally acknowledged. 

\appendix 
\section{\, Symplectic eigenvalues of the covariance matrix ${\mathcal V}^{\rm PT}$ } 

We focus on the positive definite CM ${\mathcal V}^{\rm PT}$ which is built  
with the Gaussian operator ${\hat \rho}^{\rm PT}$ obtained from the TMGS ${\hat \rho}$ 
by partial transposition of {the}  density matrix. 
Let us denote its symplectic eigenvalues  
by ${\kappa}_{\pm}^{\rm PT}$ and write down the counterparts of Eqs.\ (\ref{detV1}) and\ (\ref{D1}):
\begin{align}
\det({\mathcal V})=\left( {\kappa}_{+}^{\rm PT} \right)^2 \left( {\kappa}_{-}^{\rm PT} \right)^2, 
\label{A1}
\end{align}
\begin{align}
& {\mathcal D}^{\rm PT} =\left[ \left( {\kappa}_{+}^{\rm PT} \right)^2-\frac{1}{4} \right]
\left[ \left( {\kappa}_{-}^{\rm PT} \right)^2-\frac{1}{4} \right],
\label{A2}
\end{align}
with
\begin{align}
{\kappa}_{+}^{\rm PT} \geqq {\kappa}_{-}^{\rm PT}>0. 
\label{A3}
\end{align} 
from Eqs.\ (\ref{A1}),\ (\ref{A2}), and\ (\ref{DPT1}) we get the biquadratic equation
\begin{align}
\left( {\kappa}^{\rm PT} \right)^4-\left( b_1^2+ b_2^2-2cd \right) \left( {\kappa}^{\rm PT} \right)^2
+\det({\mathcal V})=0, 
\label{biquad}
\end{align} 
satisfied by the symplectic eigenvalues:
\begin{align}
& \left( {\kappa}_{\pm}^{\rm PT} \right)^2=\frac{1}{2}\left[ \left(b_1^2+b_2^2 -2cd \right) 
\pm \sqrt{{\Delta}^{\rm PT}} \right],      \notag \\
& {\Delta}^{\rm PT}:=\left( b_1^2+b_2^2-2cd \right)^2 -4\, \det({\mathcal V})    \notag \\ 
& =\left( b_1^2-b_2^2 \right)^2+4\left( b_1 c-b_2 d \right) \left( b_2 c-b_1 d \right) 
\geqq 0. 
\label{A5}  
\end{align}

Taking account of Eqs.\ (\ref{A2}) and\ (\ref{A3}), Simon's separability 
condition\ (\ref{DPT1}) states the following alternative:
\begin{itemize}
\item  ${\cal D}^{\rm PT} \geqq 0 \;\;  \Longleftrightarrow  \; \; 
{\kappa}_{-}^{\rm PT} \geqq \frac{1}{2}   \quad $  
for separable TMGSs;
\item  ${\cal D}^{\rm PT} < 0  \;\;  \Longleftrightarrow  \;\;  {\kappa}_{-}^{\rm PT} < \frac{1}{2}   \quad $ 
for entangled TMGSs.
\end{itemize}
Note that the partial transpose ${\hat \rho}^{\rm PT}$ of any entangled TMGS ${\hat \rho}$ 
is no longer a state.

Multiplication of Eq.\ (\ref{biquad}) by the positive quantities $b_1 b_2-c^2$ 
and $b_1 b_2-d^2$  from Eq.\ (\ref{GS2}) yields two useful identities: 
\begin{align}
& \left[ b_1 \left( b_1 b_2-c^2 \right) -b_2 \left( {\kappa}_{\pm}^{\rm PT} \right)^2 \right]
\left[ b_2 \left( b_1 b_2-c^2 \right) -b_1 \left( {\kappa}_{\pm}^{\rm PT} \right)^2 \right]     \notag \\ 
& = \left[ c \left( {\kappa}_{\pm}^{\rm PT} \right)^2 -d\left( b_1 b_2-c^2 \right) \right]^2 \, ;  
\label{A6}
\end{align}
\begin{align}
& \left[ b_1 \left( b_1 b_2-d^2 \right) -b_2 \left( {\kappa}_{\pm}^{\rm PT} \right)^2 \right]
\left[ b_2 \left( b_1 b_2-d^2 \right) -b_1 \left( {\kappa}_{\pm}^{\rm PT} \right)^2 \right]     \notag \\ 
& = \left[ c\left( b_1 b_2-d^2 \right)-d \left( {\kappa}_{\pm}^{\rm PT} \right)^2 \right]^2 . 
\label{A7}
\end{align}
Equations\ (\ref{A6}) and\ (\ref{A7}) transform into each other by interchanging
the parameters $c$ and $-d$.  Another pair of identities related in the same way 
is obtained when we multiply Eq.\ (\ref{biquad}) by the product $ b_1 b_2$:
\begin{align}
& \left[ b_1 \left( b_1 b_2-c^2 \right) -b_2 \left( {\kappa}_{\pm}^{\rm PT} \right)^2 \right]
\left[ b_2 \left( b_1 b_2-d^2 \right) -b_1 \left( {\kappa}_{\pm}^{\rm PT} \right)^2 \right]     \notag \\ 
& = \left( b_1 c-b_2 d \right)^2 \left( {\kappa}_{\pm}^{\rm PT} \right)^2 \, ;  
\label{A8} 
\end{align}  
\begin{align}
& \left[ b_1 \left( b_1 b_2-d^2 \right) -b_2 \left( {\kappa}_{\pm}^{\rm PT} \right)^2 \right]
\left[ b_2 \left( b_1 b_2-c^2 \right) -b_1 \left( {\kappa}_{\pm}^{\rm PT} \right)^2 \right]     \notag \\ 
& = \left( b_2 c-b_1 d \right)^2 \left( {\kappa}_{\pm}^{\rm PT} \right)^2 . 
\label{A9}
\end{align} 

We finally mention an identity involving the determinant\ (\ref{A2}):
\begin{align}
& 4\left[ b_2 \left( b_1 b_2-c^2 \right)-\frac{1}{4} b_1 \right] 
\left[ b_2 \left( b_1 b_2-d^2 \right) -\frac{1}{4} b_1 \right]    
\notag\\
& -\left( b_2^2 -\frac{1}{4} -c d \right)^2=4\left( b_2^2-\frac{1}{4} \right) {\mathcal D}^{\rm PT}.
\label{A10}
\end{align}

\section{Hessian matrices} 

\subsection{The function $\ln\left[ E\left( \lambda,\, \mu \right) \right]$} 

By using Eq.\ (\ref{E3}), we evaluate the second-order derivatives 
of the function $\ln{\left[ E\left( \lambda,\, \mu \right) \right] }$
at the stationary point\ (\ref{minp}):
\begin{align}
& H_{11}:=\frac{ {\partial}^2 \ln(E)}{ {\partial \lambda}^2 }
\left( {\lambda}_m,\, {\mu}_m \right),  H_{22}:=\frac{ {\partial}^2 \ln(E)}{ {\partial \mu}^2 }
\left( {\lambda}_m,\, {\mu}_m \right),    \notag \\ 
& H_{12}:=\frac{ {\partial}^2 \ln(E)}{ \partial \lambda \partial \mu} 
\left( {\lambda}_m,\, {\mu}_m \right).
\label{B1}
\end{align} 
By use of Eq.\ (\ref{stat}), we get the following entries of the Hessian matrix:
\begin{align}
& H_{11}=\frac{2\left( b_1 b_2-c^2 \right) }{ \left[ \Delta Q ({\lambda}_m) \right]^4 }>0, 
\qquad  H_{22}=\frac{2\left( b_1 b_2-d^2 \right) }{ \left[ \Delta P ({\mu}_m) \right]^4 }>0, 
 \notag \\ 
& H_{12}=-\frac{2}{\left( 1+{\lambda}_m{\mu}_m \right)^2}<0.
\label{B2}
\end{align} 
Then, in view of Eq.\ (\ref{A5}), the Hessian determinant is positive:
\begin{align}
\det(H)=\frac{4 \sqrt{ {\Delta}^{\rm PT} } }{ \left( {\kappa}_{-}^{\rm PT} \right)^2
\left( 1+{\lambda}_m{\mu}_m \right)^4}>0. 
\label{B3}
\end{align} 
Consequently, the Hessian matrix\ (\ref{B1}) is positive definite. The absolute
minimum of the function $E\left( \lambda,\, \mu \right) $,  Eq.\ (\ref{E3}), is therefore
\begin{align}
E_m=E\left( {\lambda}_m,\, {\mu}_m \right).
\label{B4}
\end{align} 

\subsection{The function $F\left( {\alpha}^2,\, u_1,\, u_2 \right)$} 

We have evaluated the Hessian matrix of the function 
$F\left( {\alpha}^2,\, u_1,\, u_2 \right)$, Eq.\ (\ref{F3}), 
at its stationary point, Eqs.\ (\ref{u_km}) and\ (\ref{a_m}). In the sequel, 
the indices $1,\, 2,\, 3\,$ refer to the independent variables $u_1,\, u_2,\, {\alpha}^2,\,$ 
respectively. Let us write the expressions of three principal minors: 
the diagonal entry $H_{11}$, the cofactor $A_{33}$, and the Hessian determinant
$\det(H)$. First, the following expression of $H_{11}$ holds provided that $c+d >0$:
\begin{align}
& H_{11}=\frac{\sqrt{\gamma} }{2\left( {\alpha}_m^2 +\frac{1}{ {\alpha}_m^2} \right)
u_{1m}^2 u_{2m} }  \notag \\
& \times \left[ \left( 1+\frac{1}{u_{1m}^2} \right) \frac{cu_{1m}u_{2m}+d}
{1-\frac{1}{u_{1m}^2} }+2\frac{c{\gamma}+d}{1-\frac{1}{u_{1m}^2}}  \right] >0.
\label{B5}
\end{align} 
A suitable simplification in Eq.\ (\ref{B5}) yields a formula which is valid also 
in the limit case of the two-mode STSs $(c+d=0):$
\begin{align}
& H_{11}=\frac{\sqrt{\gamma} }{2\left( {\alpha}_m^2 +\frac{1}{ {\alpha}_m^2} \right)
u_{1m}^2 u_{2m} }        \notag \\
& \times \frac{1}{b_1\sqrt{ {\Delta}^{\rm PT} }+b_1\left( b_1^2 -b_2^2 \right) 
-2d\left(  b_1 c-b_2 d \right) }      \notag \\
& \times \left\{ \left( 1+\frac{1}{u_{1m}^2} \right) \frac{b_1\left( b_1 b_2 -d^2 \right) }
{ {\gamma}\left( b_1 b_2 -c^2 \right) }
\left[ c\left( b_1^2 +b_2^2 \right) -d\left( 2b_1 b_2 \right)   \right. \right.   \notag \\
& \left. \left. +c\sqrt{ {\Delta}^{\rm PT} } \right]   
+4\left( b_1 b_2 -d^2 \right) \left( b_1 c-b_2 d \right) \right\} >0.
\label{B6}
\end{align} 
Second, we present a general symmetric expression of the cofactor $A_{33}:$
\begin{align}
& A_{33}=\frac{1}{\left( {\alpha}_m^2 +\frac{1}{ {\alpha}_m^2} \right)^2
u_{1m} u_{2m} }  \left\{ b_1 b_2 \left( \frac{1}{u_{2m} }-\frac{1}{u_{1m} }\right)^2       
\right.   \notag \\
& \left. + \frac{ 4\left( b_1 b_2 -c^2 \right) \left( b_1 c-b_2 d \right) \left( b_2 c-b_1 d \right) }
{ c\left( b_1^2 +b_2^2 \right) -d\left( 2b_1 b_2 \right)
+c\sqrt{ {\Delta}^{\rm PT} }  }  \right.   \notag \\
& \left. \times \frac{1}{c-d} \left( 1+\frac{1}{u_{1m} u_{2m} } \right)  \right\} >0.
\label{B7}
\end{align} 
Third, we have obtained the Hessian determinant:
\begin{align}
& \det(H)=\frac{4}{ {\alpha}_m^4 \left( {\alpha}_m^2 +\frac{1}{ {\alpha}_m^2} \right)^3
\left( u_{1m} u_{2m} \right)^{\frac{3}{2} } }    \notag \\
& \times \frac{ 4\left( b_1 b_2 -c^2 \right) \left( b_1 c-b_2 d \right) \left( b_2 c-b_1 d \right) }
{ c\left( b_1^2 +b_2^2 \right) -d\left( 2b_1 b_2 \right)+c\sqrt{ {\Delta}^{\rm PT} } } >0.
\label{B8}
\end{align} 
The obvious inequalities\ (\ref{B6})-\ (\ref{B8}) show that the Hessian matrix
under discussion is positive definite. Therefore, the unique stationary point
$\left\{  {\alpha}_m^2, u_{1m}, u_{2m} \right\} $ is a minimum point where the function 
$F\left( {\alpha}^2,\, u_1,\, u_2 \right)$ reaches its absolute minimum\ (\ref{F_m}). 

Needless to say, Eqs.\ (\ref{B6})-\ (\ref{B8}) considerably simplify in the particular cases
of STSs and symmetric states. We list the resulting formulae as follows.
\begin{enumerate}

\item {\em Two-mode STSs}:
\begin{align}
& H_{11}=\frac{1}{  \sqrt{\delta} }\left\{ b_1 \sqrt{\delta} -\left[ b_1 \left( b_1 -b_2 \right) 
+c^2 \right] \right\} >0;      \notag \\
& A_{33}=\frac{c^2}{\delta} \left( b_1 +b_2 \right) \left[ \left( b_1 +b_2 \right) 
-\sqrt{\delta} \right] >0;    \label{H>0STS1}
\end{align}

\begin{align}
& \det(H)= {\delta}^{-\frac{3}{2} } c^2 \left( b_1 +b_2 \right)     
\left[  \sqrt{\delta}+\left( b_1 -b_2 \right) \right]^2       \notag \\ 
& \times \left[ \left( b_1 +b_2 \right) -\sqrt{\delta} \right] >0.
\label{H>0STS}
\end{align}

\item {\em Symmetric TMGSs}:
\begin{align}
& H_{11}=\frac{1}{ 4}\left( \frac{b-c}{b+d} \right)^{\frac{1}{2} } \frac{1}{b+d} \left\{ b\left[ (b-c)
+(b+d) \right]   \right.   \notag \\
& \left. +2(b-c)(b+d) \right\} >0;    \notag \\
& A_{33}=\frac{1}{2} \left( \frac{b-c}{b+d} \right)^2 b\left[ (b-c)+(b+d) \right] >0;    \notag \\
& \det(H)=\left( \frac{b-c}{b+d} \right)^{\frac{3}{2} }b(b-c)(c-d) >0.
\label{H>0sym}
\end{align}
\end{enumerate}
The above formulae have been checked by direct evaluation of the corresponding 
Hessian matrices starting from Eqs.\ (\ref{STS}) and\ (\ref{sym}), respectively.

\subsection{The function $G\left( \xi,\, \eta,\, u_1\right)$} 

Let us assign the indices $1,\, 2,\, 3\,$ to the independent variables 
$u_1,\, \xi,\, \eta,\,$ respectively. The Hessian matrix of the function 
$G\left( \xi,\, \eta,\, u_1\right)$, Eq.\ (\ref{G1}), evaluated 
at the stationary point\ (\ref{mp}) has the following entries:
\begin{align}
& H_{11}=\frac{1}{\left( b_2^2 -\frac{1}{4} \right) u_{1m}^3}
\left\{ \left[ b_2 \left( b_1 b_2-d^2 \right)-\frac{1}{4}b_1 \right]   \right.  \notag \\
& \left. +\frac{1}{2}\left[ b_2 \left( b_1 b_2-c^2 \right)-\frac{1}{4}b_1 \right]  
\left( u_{1m}^2 +1 \right)  \right.  \notag \\
& \left.  +\frac{1}{2}b_2 c^2 \left( u_{1m}-1 \right)^2 
+\frac{1}{2}c\left( 2b_2 c+d \right) u_{1m}  \right\} >0,   \notag \\
& H_{22}=2b_2,     \qquad  H_{33}=2b_2,    \notag \\
& H_{12}=-c u_{1m}^{-\frac{1}{2} },  \quad   H_{13}=-d u_{1m}^{-\frac{3}{2} },   
\quad    H_{23}=-1.
\label{H}
\end{align}
Three nested principal minors are clearly positive: the diagonal entry $H_{33}$, 
the cofactor $A_{11}$, and the Hessian determinant $\det(H)$. Indeed,
\begin{align}
& H_{33}=2b_2 >0,  \qquad    A_{11}=4\left( b_2^2 -\frac{1}{4} \right) >0,    \notag \\
& \det(H)= \frac{2}{u_{1m}^3 }
\left\{ 2\left[ b_2 \left( b_1 b_2-d^2 \right)-\frac{1}{4}b_1 \right]   \right.  \notag \\
& \left. +\left[ b_2 \left( b_1 b_2-c^2 \right)-\frac{1}{4}b_1 \right]  
\left( u_{1m}^2 +1 \right)   \right.   \notag \\
& \left.  +b_2 \left( c^2 -d^2 \right) \right\} >0.
\label{Hpos}
\end{align}
Accordingly, the Hessian matrix with the entries\ (\ref{H}) is positive definite. This means 
that the unique stationary point $\left\{  {\xi}_m, {\eta}_m, u_{1m} \right\} $, 
Eq.\ (\ref{mp}), is a minimum point where the function  $G\left( \xi,\, \eta,\, u_1\right)$ 
reaches its absolute minimum\ (\ref{G_m}). 

{\bf Acknowledgments:} This work was supported by the funding agency CNCS-UEFISCDI
of the Romanian Ministry of Research and Innovation
through Grant No. PN-III-P4-ID-PCE-2016-0794.

\end{document}